\documentclass[twocolumn]{aastex631}

\usepackage{CJK}
\usepackage{soul}
\usepackage{bm}

\usepackage[normalem]{ulem}

\begin{document}
\begin{CJK*}{UTF8}{gbsn} 

\title{Asymmetry, Gap Opening and High Accretion Rate on DM Tau: \\A Hypothesis Based on Interaction of Magnetized Disk Wind with Planet}

\author[0000-0003-3728-8231]{Yinhao Wu (吴寅昊)}
\affiliation{School of Physics and Astronomy, University of Leicester, Leicester LE1 7RH, UK; \href{mailto:email@domain}{yw505@leicester.ac.uk}}

\begin{abstract}
Over two hundred protoplanetary disk systems have been resolved by ALMA, and the vast majority suggest the presence of planets. The dust gaps in transition disks are considered evidence of giant planets sculpting gas and dust under appropriate disk viscosity. However, the unusually high accretion rates in many T Tauri stars hosting transition disks challenge this theory. As the only disk currently observed with high turbulence, the high accretion rate ($\sim10^{-8.3}M_{\odot}/yr$) observed in DM Tau indicates the presence of strong turbulence may within the system. Considering the recent theoretical advancements in magnetized disk winds is challenging the traditional gap-opening theories and viscosity-driven accretion models, our study presents a pioneering simulation incorporating a simplified magnetized disk wind model to explain the observed features in DM Tau. Employing multi-fluid simulations with an embedded medium mass planet, we successfully replicate the gap formation and asymmetric structures evident in ALMA Band 6 and the recently JVLA 7 mm observations. Our results suggest that when magnetized disk wind dominate the accretion mode of the system, it's entirely possible for a planet with a medium mass to exist within the gap inside 20 au of DM Tau. This means that DM Tau may not be as turbulence as imagined. However, viscosity within the disk should also contribute a few turbulence to maintain disk stability. 

\end{abstract}


\section{Introduction} \label{sec:intro}
The Atacama Large Millimeter/submillimeter Array (ALMA) has unveiled a wide array of intricate structures within mm/sub-mm observations of protoplanetary disks (PPDs), showcasing features such as cavities, gaps, rings, and vortices, as reported in numerous studies \citep[e.g.,][]{ALMA2015,long2018,Long2019,andrews2018,Huang2018,lodato2019,Andrews2020}. Notably, the pervasive presence of luminous rings alongside dim gaps stands out. Although some recently studies suggest that substructures can be generated by specific dust distributions without relying on planet-disk interactions \citep[e.g.,][]{Pinilla_2021, Kalyaan2021, Stadler_2022, Jiang2023}. However, traditionally explained by the influence of planets embedded within the disk, altering the distributions of gas and dust \citep[e.g.,][]{PaardekooperEtal2022}. 

The interaction of giant planets with the disk leads to the generation and attenuation of density waves at resonances, creating noticeable depletions in the gas density nearby \citep{Goldreich_Tremaine_1980, Lin_Papaloizou1986a, Lin_Papaloizou_1993}. This interaction results in the formation of a pressure bump at the gas gap's outer boundary, effectively halting the inward migration of dust particles poorly coupled to the gas \citep{paardekooper2006,Zhuetal2011}, hence the formation of pronounced dust rings. By adjusting parameters such as disk viscosity, thickness, and planet mass, one can replicate an assortment of dust emission patterns observed in the sub-mm ALMA observations \citep[e.g.,][]{Rosotti2016,2018ZhangDSHARP,Chen2021,Garrido-Deutelmoser_2023_HD163296,Wu_2023_migration}.

In the process of planet-disk interaction, the turbulent viscosity parameter $\alpha_{\rm v}$ \citep{Shakura-Sunyaev73} plays a role in maintaining the disk profile and preventing gap opening. However, current observational results suggest that the planet-forming regions in PPDs are not as turbulent as previously thought \citep{Flaherty2017,Lodato2017,Dullemond2018,Rosotti2020,DoiKataoka2021}. Although the methods used for these observations may vary, and moreover,  some gas observations trace the upper disk layer in the outer parts of the disk. These latest observations can restrict the upper limit of the disk's turbulence coefficient to $10^{-4} \sim 10^{-3}$, meaning that if PPDs are indeed weak turbulence or nearly inviscid, then a viable alternative explanation is needed for how disks accrete and evolve. 

Recent theoretical work proposes that magnetically-driven disk winds might transport the angular momentum of gas at a faster rate than disk turbulence, dominating the accretion in PPDs \citep{Bai2011,Bai2013,Bai2015,Armitage2013,2013BaiStone,Bai2014,Suzuki2016,Bai2016,Hasegawa2017,Lesur2021,Cui2021,Cui2022}. This could be a promising alternative to reconcile the relationship between turbulence velocity and host stellar accretion rates. 

Based on numerical simulation of planet-disk interactions in with MHD disk winds, \cite{MHD-wind-Elbakyan} and \cite{Aoyama-Bai-2023} have shown that for Type-II planets, gap opening is easier under the same level of turbulence when magnetohydrodynamics (MHD) disk wind dominates accretion, compared to the traditional viscosity-driven accretion mode. And \cite{WU_Chen_jiang_2023} has further discovered that when MHD disk wind dominates in the angular momentum transfer within the disk, different degrees of asymmetric structures can be generated depending on the strength of the wind. These structures, distinct from those produced in viscosity-dominated disks \citep{2018ZhangDSHARP,Long2022LkCa15Disk}, can be directly imaged by ALMA.

In this work, we use a simplified MHD disk wind prescription for the first time to simulate a case study of DM Tau—the well studied transition disk around the T Tauri star \citep[e.g.,][]{Chung2024}. We focus on reproduce its substructures, and keep it with a high accretion rate. By embedding a medium mass ($\sim53$ times Earth mass) planet and using multi-fluid in simulations, we reproduce the gap opening and asymmetric structures caused by planet-disk interaction, and make synthetic images to compare with ALMA Band 6 (230 GHz) and Karl G. Jansky Very Large Array (JVLA) Q Band (44 GHz) observations. It need be noticed that the hypothesis of multi-planet cannot be ruled out in DM Tau system \citep{Xu-Wang-2024}. However, for the sake of simplicity, we only consider the case of single-planet in this exploratory work.

The paper is organized as follows: In Sec.\ref{sec: dmtau}, we introduce the characteristics of DM Tau. In Sec.\ref{sec: simulation}, we introduce the numerical setup and parameters for implementing hydrodynamic simulations and making synthetic maps. In Sec.\ref{sec: results}, we describe the comparison between simulated observational images obtained using the MHD disk wind model and actual ALMA and JVLA observations. The discussions and conclusions are detailed in Sec.\ref{sec: discussions} and Sec.\ref{sec: conclusions}.

\section{The DM Tau Disk System}\label{sec: dmtau}
DM Tau is a T Tauri star with a high active accretion level $\sim10^{-8.3}M_{\odot}/yr$, located at 145 pc far from us with a mass of 0.53 $M_{\odot}$ \citep{Gaia_2023}. It hosts a transition disk inclined at i $\sim$ $36.1^{\circ}$ and extending to 120 au in dust emission, which radius encloses 90\% of the dust disk emission. 

Recent observational studies have confirmed the dim gap at approximately 4 to 21 au from the center of DM Tau, with a bright ring at 21 au that features some asymmetric clumpy structures \citep{Kudo2018,Hashimoto2021}. This suggests that there might be forming planets within the 4-21 au region. This type of structure, also observed in some other disks. In general, it might be generated by a super-Earth-mass planet residing in a system with low viscosity \citep{Dong_2017_super-earth,Dong_2018_super-earth,Perez_2019,Francis_Nienke_2020}.

However, if the high accretion rate implied by observations is generated by MRI viscosity, it would mean that this source is not a low-viscosity but a high-viscosity system \citep{Delage_2022_MRI_Viscosity}. This would require a sufficiently massive planet (may be nearly 10 times the mass of Jupiter) to produce such structures \citep{Pinilla_2012,Francis2022}. And the massive planets will carve out eccentric gaps and generate streamer structures \citep{ChenKan2024}, thus the hypothesis of a single-massive-planet would have difficulty generating the observed asymmetric structures in DM Tau.

At the same time, recent observations have shown that the two asymmetric structures on the DM Tau ring exhibit different observational characteristics at different wavelengths. Specifically, in previous high-resolution ALMA Band 6 observations \citep{Kudo2018,Hashimoto2021}, the asymmetric structure on the west side of DM Tau was brighter than the one on the south side. However, in the latest JVLA 7 mm observations \citep{Baobab_2024}, the asymmetric structure on the south side of DM Tau is much brighter, showing a major asymmetry, while the west side asymmetric structure has become very dim. These observational results suggest that the formation of substructures in DM Tau requires a new theoretical explanation.

\section{Numerical Setup}\label{sec: simulation}
To use MHD disk wind model to fit the observable signatures on DM Tau, we use the hydrodynamic grid-basics code \texttt{FARGO3D} \citep{FARGO3D} with its multi-fluid version \citep{FARGO3D-multifluid}.  

\subsection{Hydrodynamic Model}\label{subsec:hydro}
Our work consider a 2D cylindrical polar coordinate ($r$, $\varphi$), locally isothermal protoplanetary disk system with an embedded planet, composed of gas and dust. We neglect the disk self-gravity, planet migration and accretion in all of our simulations. The disk model in our simulations is described by the Navier-Stokes equations:
\begin{equation}
    \frac{\partial\Sigma_{\rm g}}{\partial t}+\nabla\cdot\left(\Sigma_{\rm g}\textbf{\emph{V}}\right)=0\;,
    \label{HD-dens-gas}
\end{equation}
\begin{equation}
    \frac{\partial\boldsymbol{V}}{\partial t}+\left(\boldsymbol{V}\cdot\nabla\right)\boldsymbol{V}=-\nabla\Phi-\frac{\nabla P}{\Sigma_{\rm g}}-\frac{\sum\limits_{i}^{} \Sigma_{\rm d, \it i}\boldsymbol{F}_i}{\Sigma_{\rm g}}-\nabla\cdot\tau + \boldsymbol{S}\;,
    \label{HD-vel-gas}
\end{equation}

\begin{equation}
    \frac{\partial\Sigma_{\rm d, \it i}}{\partial t}+\nabla\cdot\left(\Sigma_{\rm g}\boldsymbol{W}_i+\boldsymbol{j_i}\right)=0\;,
    \label{HD-dens-dust}
\end{equation}
\begin{equation}
    \frac{\partial\boldsymbol{W}_i}{\partial t}+\left(\boldsymbol{W}_i\cdot\nabla\right)\boldsymbol{W}_i=-\nabla\Phi+\boldsymbol{F}_i\;,
    \label{HD-vel-dust}
\end{equation}
where $\Sigma_{\rm g}$, $\textbf{\emph{V}}$ and $P$ are the surface density, velocity and pressure of the gas, respectively. The $\Phi$ is the gravitational potential from the central stellar and the planet. The $\Sigma_{\rm d, \it i}$, $\boldsymbol{W}_i$, $\boldsymbol{j_i}$ and $\boldsymbol{F}_i$ is the surface density, velocity, dust diffusion flux and the drag force from gas acting on the dust of unit mass for $i$-th dust species, respectively. And the $\tau$ is viscous stress tensor.

In general, dust is turbulently diffused in viscous disks, with its mass flux following the gradient of gas density to replenish the gap. However, in MHD windy disks, the wind solely extracts angular momentum, leading to a consistently inward gas velocity. Based on the prescription given by \citet{MHD-wind-Elbakyan}, which is an extension from 1D prescription \citet{Tabone22,Tabone2022-2}, we introduced the wind accretion term $\boldsymbol{S}$ in Eq. \ref{HD-vel-gas}, to simulate the role of the MHD disk wind in the angular momentum transfer process within the disk:
\begin{equation}
    \boldsymbol{S} = \Gamma \frac{\hat{\boldsymbol{r}}}{r},
\end{equation}
here ${\hat{\boldsymbol{r}}}$ is set to the unit vector in radial direction. The $\Gamma$ is a distinct torque, that serves to reduce the rotational angular momentum of the gas:
\begin{equation}
    \Gamma = \frac{1}{2} \sqrt{ \frac{GM_*}{r}} V_{\rm dw}\;.
    \label{Gamma}
\end{equation}
For DM Tau, we set $M_{\star} = 0.53~M_{\odot}$ of the host star in our simulations \citep{Gaia_2023}, and $V_{\rm dw}$ is the characteristic radial inward velocity of gas due to MHD disk winds corresponding to this torque follow \citet{Tabone22}:
\begin{equation}
        V_{\rm dw} = -\frac{3}{2}\alpha_{\rm dw} h^2 V_{\rm K}\;,
    \label{v_dw}
\end{equation}
here $\alpha_{\rm dw}$ is a dimensionless parameter to describe the strength of the disk wind, defined in a similar manner to the turbulent viscosity parameter $\alpha_{\rm v}$. Therefore, the total accretion level in the disk is $\alpha_{\rm acc} = \alpha_{\rm dw} + \alpha_{\rm v}$, see \cite{Tabone22} for more details. And $h$ is the disk aspect ratio, $V_{\rm K}$ is the Keplerian velocity at $r$.

To summarize, the impact of wind torque (Eq.\ref{Gamma}) translates to an augmented radial velocity in the gas (Eq.\ref{v_dw}), consistently directed towards the host star. This wind-induced radial gas motion can match or exceed the viscous velocity of the gas, contingent on the magnitude of $\alpha_{\rm dw}$. This effect, in turn, significantly affects the radial drift velocity of dust particles, facilitating the transport of pebbles across any gap \citep{WU_Chen_jiang_2023}.

\subsection{Numerical Parameters}\label{subsec:parameters}
The simulation spans a computational domain from 0.3 $r_0$ to 5.0 $r_0$, with $r_0 =20$ au denoting the unit length in code, the planet is also fixed at this radial position. The domain resolution includes 512 radial cells and 656 azimuthal cells. This resolution comes from previous simulations of 2D MHD disk winds using \texttt{FARGO3D} \citep{Kimmig-2020,MHD-wind-Elbakyan,WU_Chen_jiang_2023}.

We set the initial gas surface density profile to
\begin{equation}
    \Sigma_{\rm g}(r) = \Sigma_0 ({r}/{r_0})^{-1}\;,
    \label{sigma-disc0}
\end{equation}
where $\Sigma_0 = 4 \times 10^{-4} M_{\star}/r_0^2$ in code units, this means that the disk mass is about 0.005 $M_{\odot}$ within a range of $\sim$100 au, which is a little lower than the value of recently estimates for DM Tau \citep{Trapman_2022}, given by a novel way using N$_{2}$H$^{+}$ and C$^{18}$O. 

In our model featuring a local isothermal temperature distribution, the aspect ratio is determined as follows:
\begin{equation}
h(r) = c_s/V_{\rm K} = H/r = h_{0}(r/r_0)^{0.25},
\end{equation}
here we set $h_{0}=0.03$, which means a flatter disk compared to some other disks is considered in our simulations.

In our model, we introduce five distinct dust populations, with particle sizes ranging logarithmically and uniformly from 30 $\mu$m to 3 mm. The dust feedback is included in our simulations. The size distribution at the onset follows a power law of $s^{-3.5}$, as suggested by \citep{MRN}. We set the initial dust-to-gas mass ratio to $\epsilon=1\%$, and assume a uniform internal density of $2.3~g/cm^{3}$ for the dust grains. 

In terms of boundary conditions, we implement wave-damping boundary conditions for the gas component, where the gas density and velocity are gradually adjusted back to their original radial profiles, to maintain simulation stability. For the dust components, we employ open boundary conditions at the inner edge of the grid, specifically for their radial velocities. This is to prevent dust from gathering on the inner region, thus avoiding the influence of boundary effects on subsequent radiation transfer processing and simobserve. 

\subsection{Simobserve}\label{subsec:observe}
We utilized the publicly software package \texttt{RADMC-3D} code \citep{RADMC-3D} to conduct our dust radiative transfer calculations. Then to process our \texttt{FARGO3D} simulation results for use as inputs in \texttt{RADMC-3D}, we used the publicly available code \texttt{fargo2radmc3d} \citep{fargo2radmc3d-dust}. The physical details of the radiative transfer calculations are described in the Appendix \ref{appendix: RT}.

To calculate the dust absorption and scattering opacities, we apply Mie theory, utilizing \texttt{OpTool} \citep{optool-2021} for the computations. The dust composition follows that used in the DSHARP project \citep{DSHARP-V,2018ZhangDSHARP}, encompassing both absorption and scattering total opacities for specified wavelengths and grain sizes. For the determination of the dust continuum emission's specific intensity, we deploy ray tracing with $10^8$ photons for scattering.

Based on the known observation data of DM Tau, we assume the disk distance is 145 parsecs far from us \citep{DMTau145pc_2007,Gaia2021,Gaia_2023}. And the star radius is set to 0.60 solar radii with a stellar effective temperature of 3705 K \citep{Gaia_2023}. For the geometry of the disk, we set the disk inclination of DM Tau is $35.2^{\circ}$ and the disk position angle is $158^{\circ}$ \citep{Kudo2018}.

To mock ALMA and JVLA observations, we used the \texttt{SIMOBSERVE} and \texttt{SIMANALYZE} functions within the \texttt{CASA 6.4.9} package \citep{CASA2007}, converting synthetic images into Fourier domain visibility data sets. We utilized the alma.out28.cfg and vla.a.cfg files provided within the \texttt{CASA} package, which include the longest baselines for ALMA (extending up to 16 km) and JVLA (extending up to 21 km). We adjusted the ``inbright" and ``integration" parameter to make The root-mean-square (RMS) noise level produced by \texttt{CASA} as close as possible to the values generated by realistic observations \citep{Hashimoto2021,Baobab_2024}.

\section{Results}\label{sec: results}
\cite{WU_Chen_jiang_2023} explored a wide range of $\alpha_{\rm v}-\alpha_{\rm dw}$ parameter space based on a fixed planet mass. In this exploratory work, we use the same planet-to-stellar mass ratio $q=0.0003$ as in that parameter space. Which corresponds to approximately 53 times Earth mass in DM Tau. As introduced in Sec.\ref{subsec:parameters}, we placed this planet in a fixed orbit at a distance of 20 au from the stellar. 

We note that some recent studies \citep[e.g.,][]{McNally_2019,Wu_2023_migration} suggest that planet migration may lead to the formation of different substructures within the disk, and the migration behavior of planets in a windy-disk differs from that in a traditional laminar disk \citep{Kimmig-2020}. However, we have overlooked the impact of planet migration in this work like most similar works in the past \citep[e.g.,][]{2018ZhangDSHARP,Ricci2018,WU_Chen_jiang_2023,Wu_2024_ska_ngvla}. In next work, we will explore planet migration within MHD disk wind (Wu \& Chen 2024 in prep). 

All of our simulations were run for 1500 orbital periods at $r_0$ (corresponding to $\sim 0.2$ Myr for physical units), which is a reasonable value for the age of DM Tau \citep{Simon2000,Luhman2010,Guilloteau2014}. 

\begin{figure*}
\centering
\includegraphics[width=0.49\hsize]{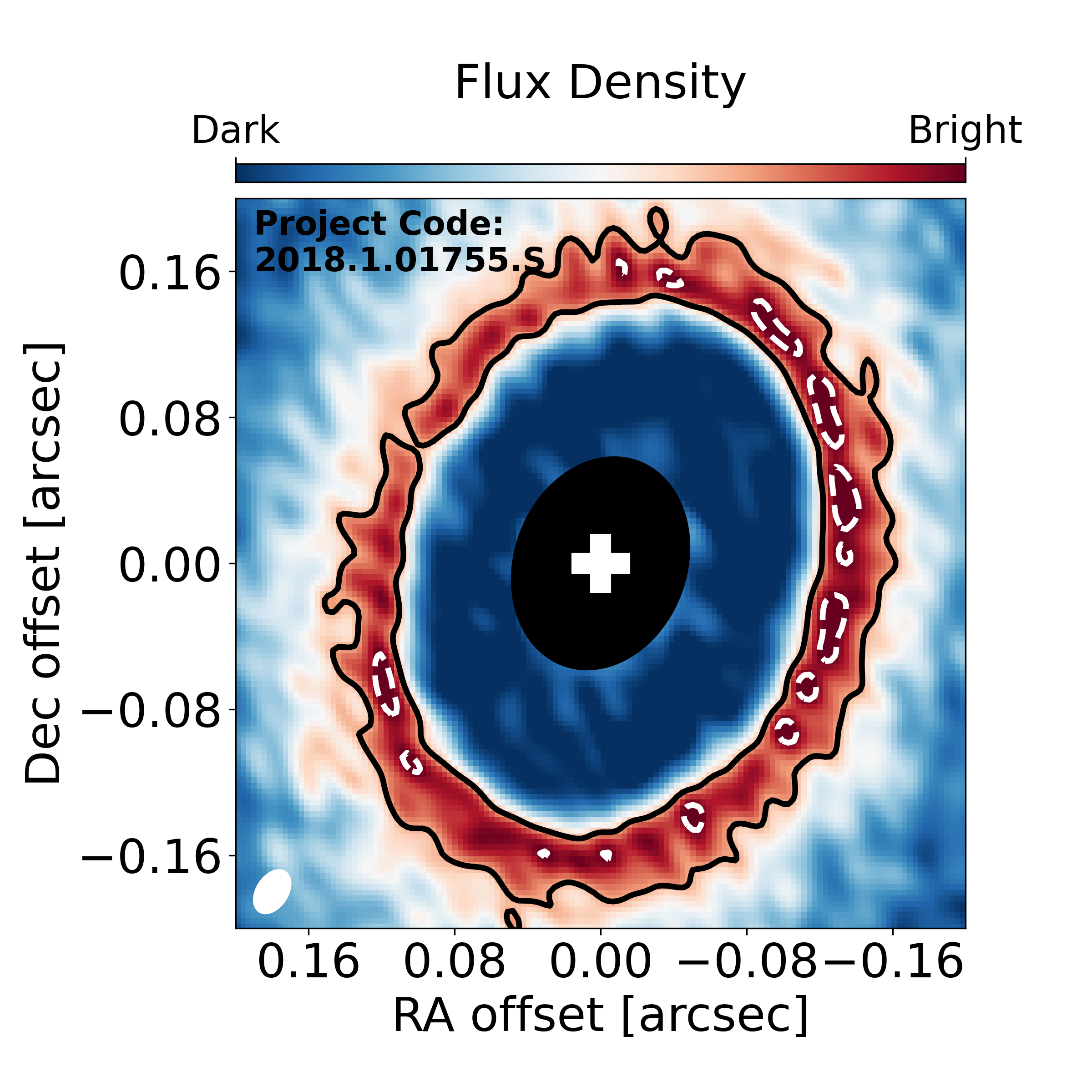}
\includegraphics[width=0.49\hsize]{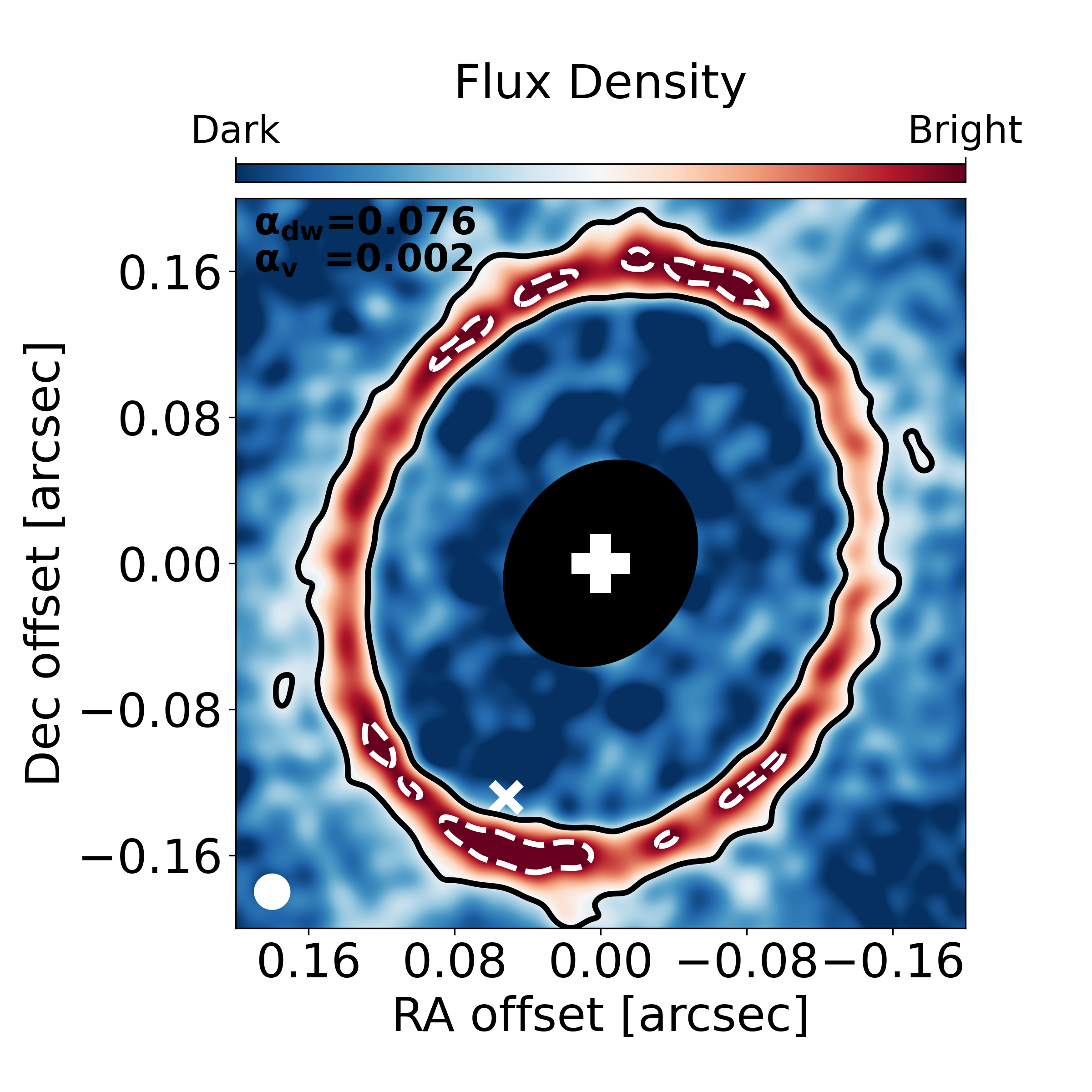}
\caption{\textit{Left panel:} Previously published ALMA Band 6 (230 GHz, $\lambda=1.3$ mm) continuum image on DM Tau \citep{Hashimoto2021} from ALMA data $\#$2018.1.01755.S (the project code marked at the upper-left corner), with the $0.027\times0.018$ arcsec beam size (beam position angle is $33^{\circ}$). \textit{Right panel:} Simulated ALMA Band 6 continuum images of our fiducial model (the values of $\alpha_{\rm dw}$ and $\alpha_{\rm v}$ are written in the upper-left corner) with a 53 times Earth mass planet at 20 au (marked by white ``x") from the central star, with the $0.02\times0.02$ arcsec beam size. We mark the beam size at the left-lower corner for each panel.  The RMS noise in each panel is at the same level, i.e. RMS = $8.4\mu Jy/beam$. Since the inner boundary of our simulation domain is at 6 au, to prevent the influence of boundary effects, we've applied a mask with a black ellipse to each panel. Additionally, we've marked the position of the stellar with a white ``+" for reference. In each panel, the approximate outline of the bright ring around 21 au is delineated with black solid line contours, and the high-brightness clumps on the ring, representing the asymmetric structures, are indicated with white dashed line contours. }
\label{fig:best-fit}
\end{figure*}

\begin{figure*}
\centering
\includegraphics[width=0.49\hsize]{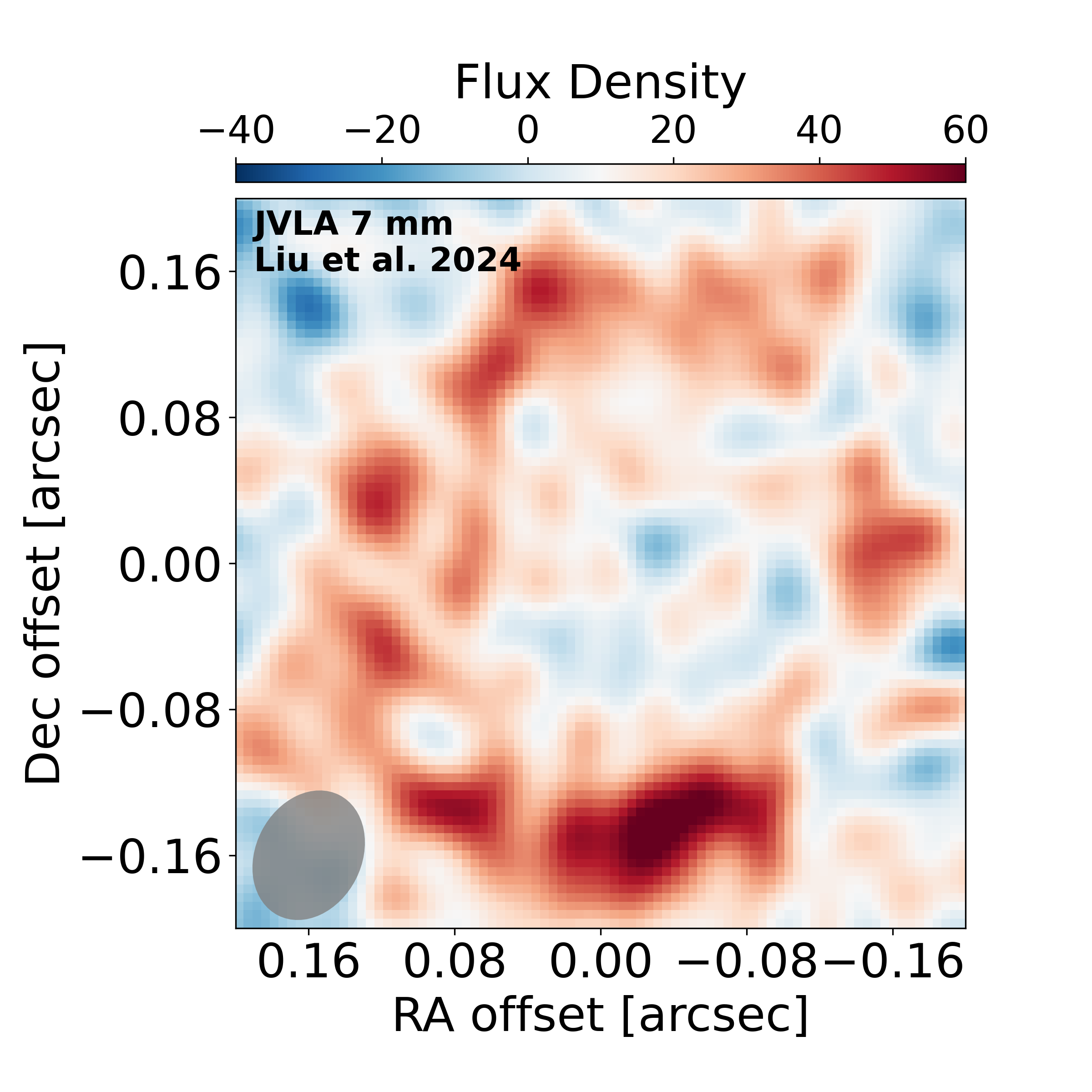}
\includegraphics[width=0.49\hsize]{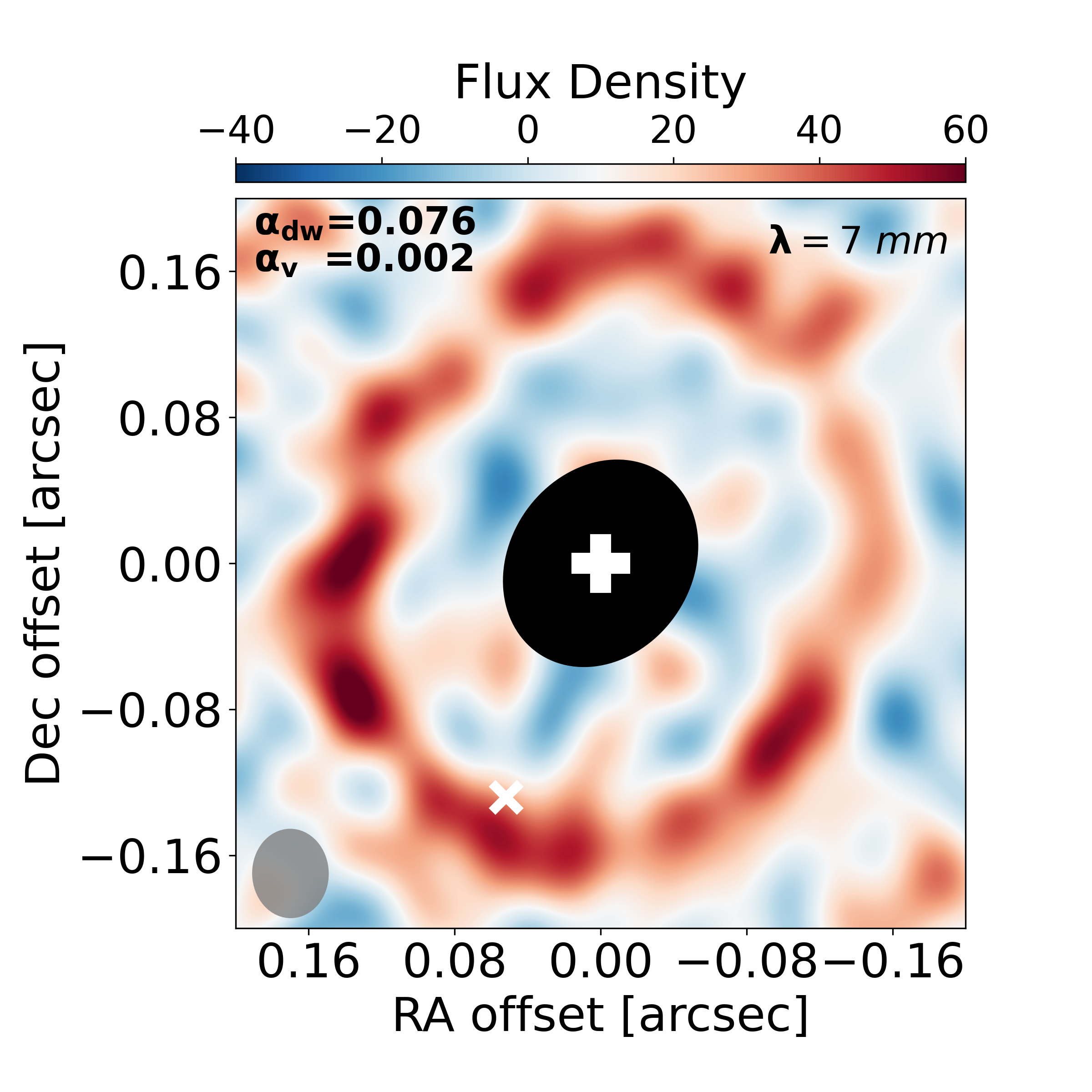}
\caption{Similar to the two top panels in Fig.\ref{fig:best-fit}, but for JVLA Q Band (40$\sim$48 GHz, $\lambda=7$ mm) continuum image \citep[][the left panel]{Baobab_2024}, and simulated JVLA continuum image (the right panel) with our fiducial model.}
\label{fig:JVLA}
\end{figure*}

Combining the parameter space study from \cite{WU_Chen_jiang_2023} with the morphological characteristics of DM Tau in dust continuum emission, we conclude that the roles of MHD disk wind and viscosity in the angular momentum transfer process of DM Tau should be well-matched in strength, i.e. $\alpha_{\rm v}\sim h\alpha_{\rm dw}$ \citep[see Figure 2 in ][]{WU_Chen_jiang_2023}. Based this, we obtained our fiducial model (as shown in Fig.\ref{fig:best-fit} and Fig.\ref{fig:JVLA}) by adjusting $\alpha_{\rm dw}$ and $\alpha_{\rm v}$. The numerical parameters used in the fiducial model are listed in Tab.\ref{table:1}. 

\begin{deluxetable}{lc}
  \tablecaption{The parameters used in fiducial model for DM Tau. Where $r_{p}$ and $M_{p}$ are the planet position and planet mass, respectively. $M_{\oplus}$ is the Earth mass and planet to stellar mass ratio is $q$.
\label{table:1}}
    \tablehead{
    \colhead{Parameter} & \colhead{Value}
    } 
\startdata
\hline\hline
$r_{p}$ & 20 au \\
$M_{p}$ & 53 $M_{\oplus}$ \\
$M_{\star}$ & 0.53 $M_{\odot}$ \\
$q$ & $3\times10^{-4}$ \\
$\Sigma_{0} $ & $4.7g/cm^{2}$ \\
$h_{0}$ & 0.03 \\
$\alpha_{\rm dw}$ & 0.076 ($\sim97\%$ of $\alpha_{\rm acc}$) \\
$\alpha_{\rm v}$ & $0.002$ ($\sim3\%$ of $\alpha_{\rm acc}$) \\
\enddata
\end{deluxetable}

\subsection{Observation at 1.3 mm and 7 mm}\label{subsec:ALMA and JVLA}
As shown in Fig.\ref{fig:best-fit}, when $\alpha_{\rm dw}=0.076$ and $\alpha_{\rm v}=0.002$, our fiducial model matches the observations very well. At this point, the MHD disk wind contributes $\sim97\%$ of the accretion rate in the system, while viscosity accounts for the remaining $\sim3\%$. Under this scenario, some of the characteristic substructures reflected in ALMA observations of DM Tau, the uneven bright ring at 21 au, the high-brightness clumps on the ring causing asymmetric structures, and the wide and deep gap spanning several au, are all represented in our fiducial model.

What is even more remarkable is that our fiducial model continues to show a disk morphology that matches the observed results at longer wavelengths. As shown in Fig.\ref{fig:JVLA}, the high-contrast clump on the ring observed in the recently JVLA 7 mm observations \citep{Baobab_2024}, which causes asymmetry, is also represented in our fiducial model. It also accurately reproduces the uneven distribution of emission on DM Tau's dust ring. The situation where some clumps with high contrast at 1.3 mm observations appear dimmer in the 7 mm observations is also reflected in the fiducial model.

The substructures in the disk displayed in the fiducial model originate from planet-disk interactions as well as the roles of wind-viscosity in the angular momentum transfer process. We will provide a brief discussion based on the simulation results in Sec.\ref{subsec:discuss 2}.

However, we noted that compared to our simulation, the latest 7 mm observations show the gap within 20 au of DM Tau to be a bit more shallow. The reason for this observation discrepancy is that our radiative transfer calculations at 1.3 mm and 7 mm wavelengths considered only dust thermal emission. Yet, \cite{Baobab_2024} concluded that at 7 mm wavelengths for DM Tau, flux densities are contributed by both dust thermal emission and time-varying free-free emission of gas \citep[e.g.,][]{Zapata2017,Liu2021}. The differences between observations and simulations in this regard precisely validate this conclusions about varying emissions contributed at different wavelengths in DM Tau.

We also remind readers that due to a lack of detailed parameters for DM Tau, our fiducial model currently cannot produce synthetic images at 1.3 mm wavelength that match the actual observations. This might be because we currently do not have a sufficient understanding of the dust species composition in DM Tau. This will require enhancements based on further observational data and analysis of DM Tau in the future.

\subsection{Pure Wind or Pure Viscosity Cases}\label{subsec:reference cases}
\begin{figure*}
\centering
\includegraphics[width=0.49\hsize]{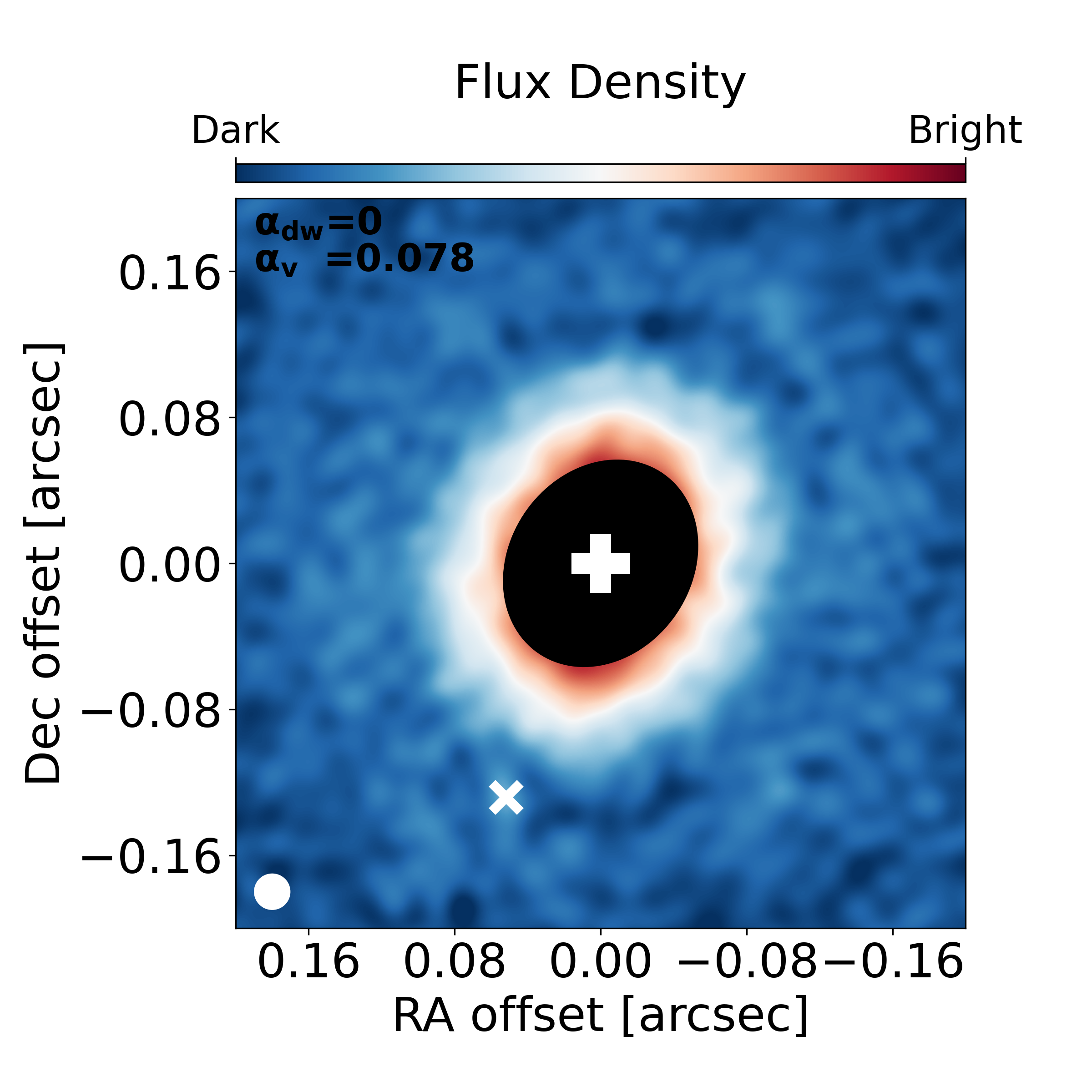}
\includegraphics[width=0.49\hsize]{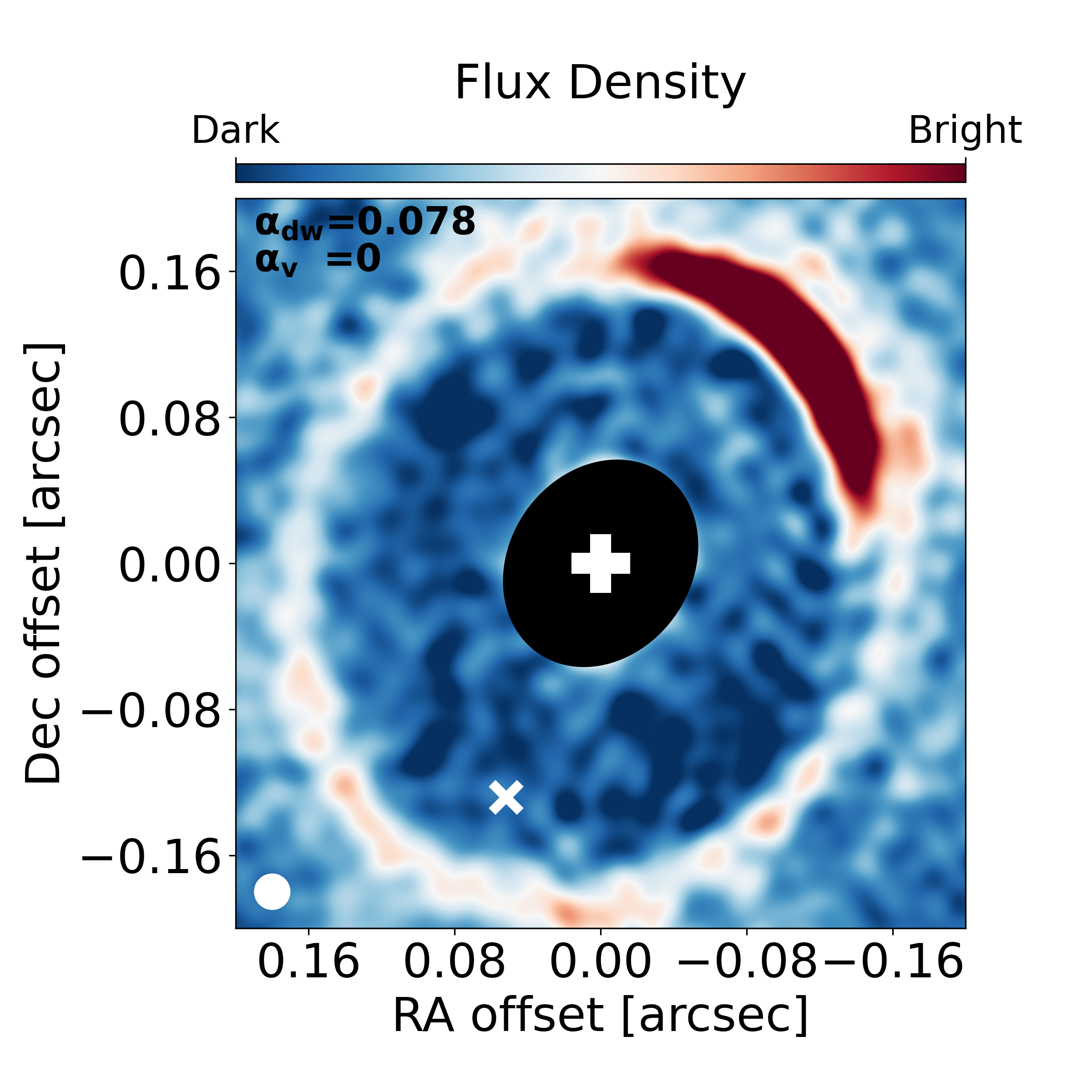}
\includegraphics[width=0.49\hsize]{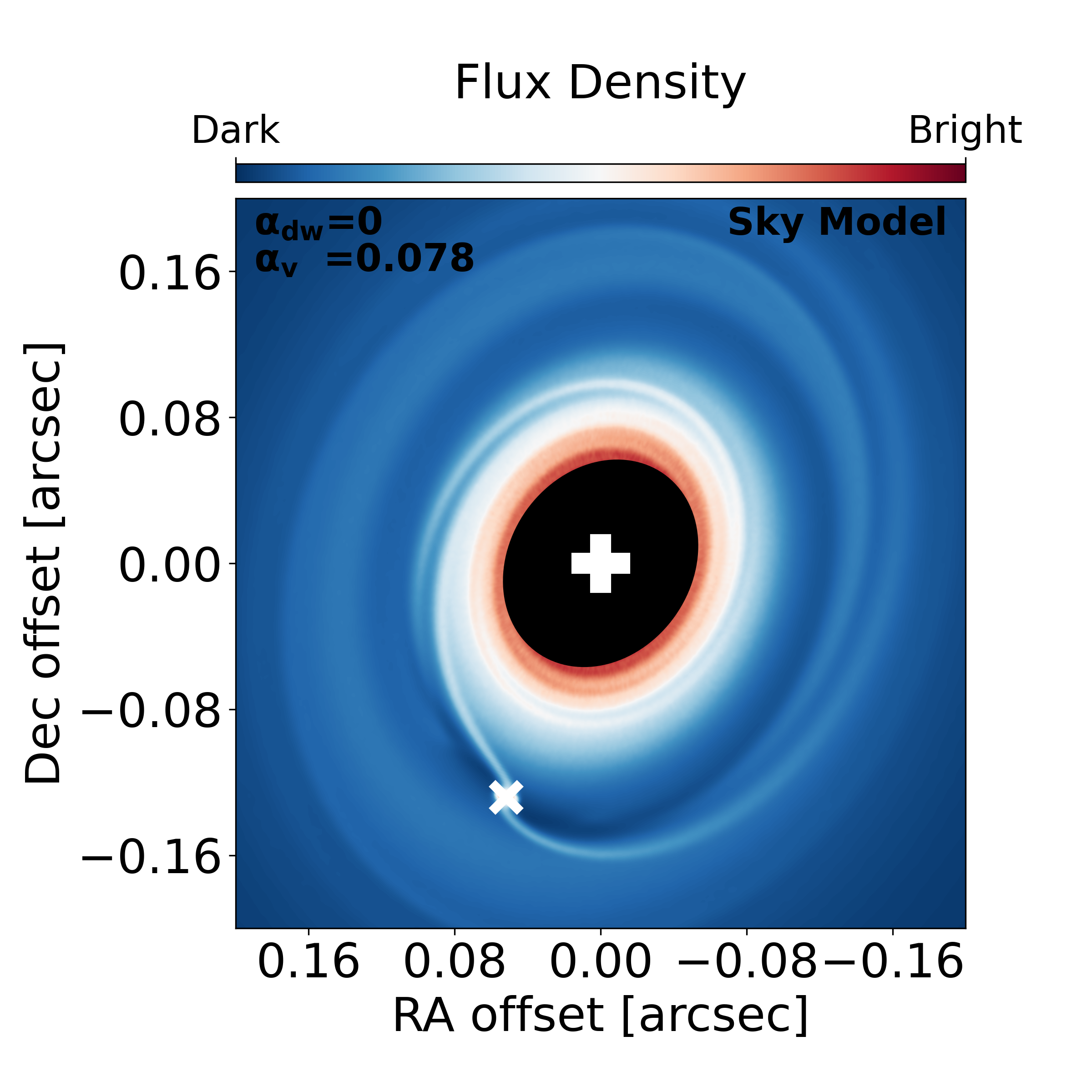}
\includegraphics[width=0.49\hsize]{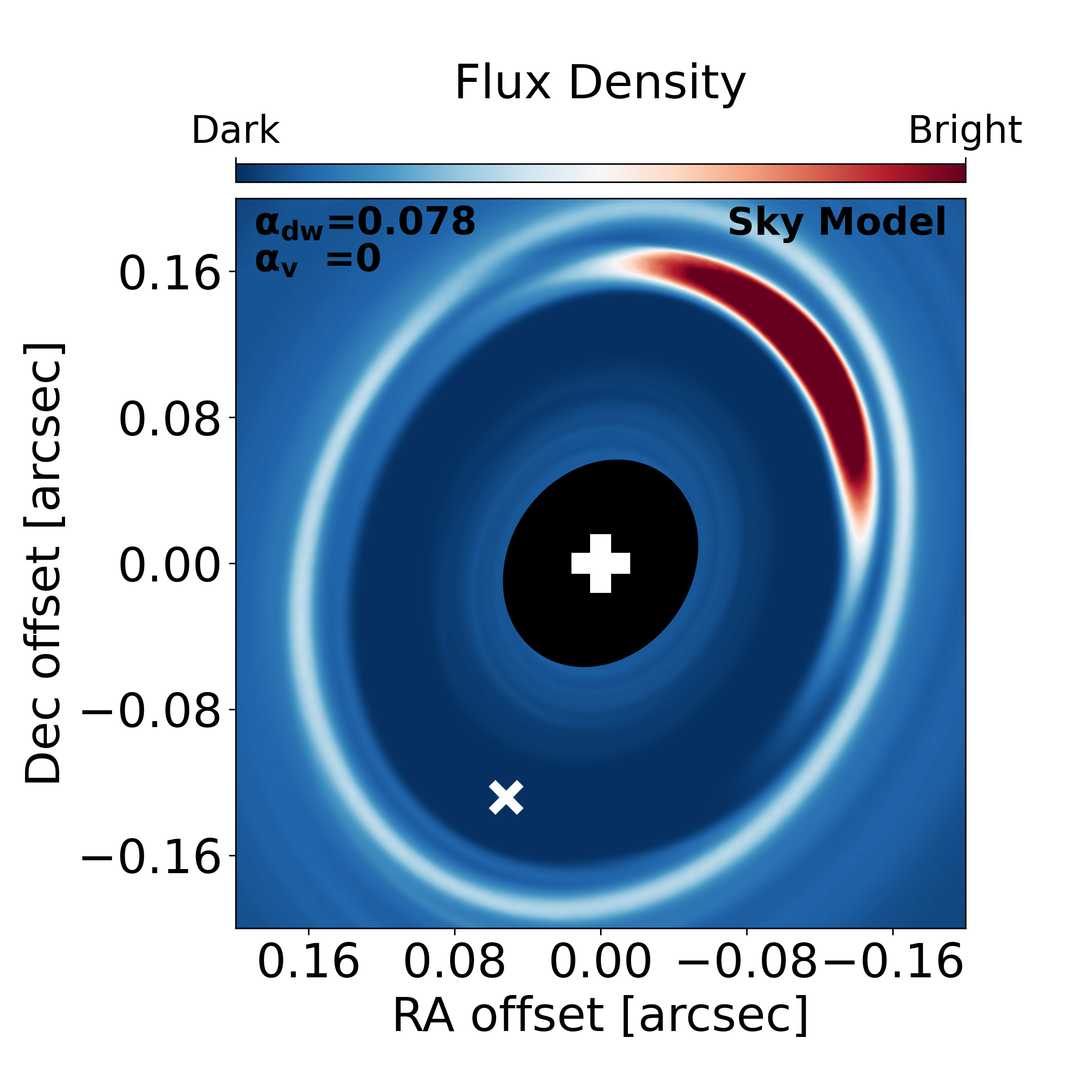}
\caption{\textit{Top panels:} Similar to Fig.\ref{fig:best-fit}, but for simulated ALMA continuum images with pure viscosity case (the left panel) and pure wind case (the right panel). \textit{Lower panels:} The sky model processed by \texttt{RADMC-3D} for reference cases.}
\label{fig:reference case}
\end{figure*}

We also plotted pure viscosity cases and pure MHD disk wind cases as reference models. As shown in Fig.\ref{fig:reference case}, when $\alpha_{\rm v}=\alpha=0.078$, a planet with 53 $M_{\oplus}$ is unable to carve out a gap that is sufficiently deep and wide in the disk to allow for the formation of a bright dust ring behind the planet. Conversely, when $\alpha_{\rm dw}=\alpha=0.078$, the strong wind interacting with the planet causes dust to accumulate on the ring in a crescent-shaped structure, creating a very pronounced and intense asymmetric structure that contrasts sharply with the surrounding ring, which does not match current observations.

The results of the reference models indicate that the accretion in DM Tau must stem from the interaction between wind and viscosity, where the wind contributes the majority of the accretion rate ($\sim97\%$) to achieve the formation of asymmetric structures and gap opening. Although the contribution of viscosity to accretion is only a small fraction ($\sim3\%$), it plays an important role in maintaining the stability of the disk structure. 

\section{Discussions}\label{sec: discussions}

Compared multi-fluid simulations to the latest ALMA Band 6 and JVLA Q Band observations, we found that the accretion in DM Tau should stem from the combined action of MHD disk wind and viscosity. In terms of contribution to turbulence, MHD disk wind will dominate. Yet in shaping the morphology and substructures of the disk, the roles of MHD disk wind and viscosity are equally important, despite viscosity contributing only a minor part to the accretion rate.

For regions within ~30 au of PPDs, this possibility exists. Through 2D MHD simulations in spherical polar coordinates with the Hall effect and ambipolar diffusion, \cite{Bai2017} found that when a poloidal field is aligned with disk rotation, the contributions of viscosity and wind to angular momentum transfer in the midplane region are comparable.

The substructures within DM Tau should also depend on planet-disk interaction and do not require planets of larger than Jupiter. Observations suggest that within the 20 au region of DM Tau \citep{Flaherty2020}, there exists sufficient material for the formation of planets up to 100 times the mass of Earth. Our study believes that planets of this mass are already capable of producing substructures consistent with observations.

Without considering the complexities of multi-planet systems \citep{Xu-Wang-2024}, the currently known models show that only the MHD disk wind model can simultaneously explain the asymmetry, gap opening, and high accretion rate observed in DM Tau. Other models, such as the gravitational instability model \citep{Xu2022,Xu2023}, can reproduce asymmetric clumps and high accretion rates but fail to account for the gap opening issue. And the dusty Rossby wave instability \citep{Liu-Bai-2023} can explain the asymmetry and gap opening in the ring but fails to account for the high accretion rate observed in DM Tau.

\subsection{Gap Opening}\label{subsec:discuss 1}
Traditionally, the required minimum gap opening planet to stellar mass ratio is very sensitive to disk aspect ratio $H/r$ and viscosity $\alpha_{\rm v}$. \cite{Crida2006} defined a parameter $\mathcal{P}$ 
\begin{equation}
    \mathcal{P} = \frac{3H}{4r_{\rm H}} + \frac{50 \alpha_{\rm v}}{q} \left({\frac{H}{r}}\right)^2 
\label{eq:CridaP0}
\end{equation}
to quantify the relationship between gap opening and planet to stellar mass ratio. At $\mathcal{P} \textless 1$ the planet migrates in the type II regime, with the potential to open a gap with a factor of 10 depression in the gas surface density. Here $r_{\rm H} = r (M_{\rm p}/3M_{\star})^{1/3}$ is the Hill radius.

When the contribution of MHD disk wind to turbulence is taken into account, \cite{MHD-wind-Elbakyan} derived Eq.\ref{eq:CridaP0} to
\begin{equation}
    \mathcal{P} =  \frac{3H}{4r_{\rm H}} + \frac{50 \alpha_{\rm v}}{q} \left({\frac{H}{r}}\right)^2  + \frac{70\alpha_{\rm dw}^{3/2}}{q} \left({\frac{H}{r}}\right)^3.
\end{equation}
Obviously, when $\alpha_{\rm dw}\gg\alpha_{\rm v}$ , the value of $\mathcal{P}$ will be dominated by $\alpha_{\rm dw}$. This also means that the extent of gap opening at this time will be determined by the strength of the MHD disk wind.

\subsection{Asymmetry in Ring}\label{subsec:discuss 2}
According to the toy model which given by \cite{WU_Chen_jiang_2023}, the asymmetric structures on the ring should originate from vortices. One popular hypothesis is that vortices arise due to the Rossby Wave Instability \citep[RWI,][]{RWI-1999,rwi-2001}, triggered by sharp gradients in the gas surface density at the boundaries of planetary gaps or other perturbation. These vortices can capture dust into clumps, creating asymmetries observable in brightness maps. Typically, sharper gas surface density gradients promote RWI development. Although both wind and viscosity work to reduce the contrast in gas surface density across a planetary gap \citep{MHD-wind-Elbakyan}, their impact on sustaining vortices appears to be contradictory.

\begin{figure}
\centering
\includegraphics[width=1\hsize]{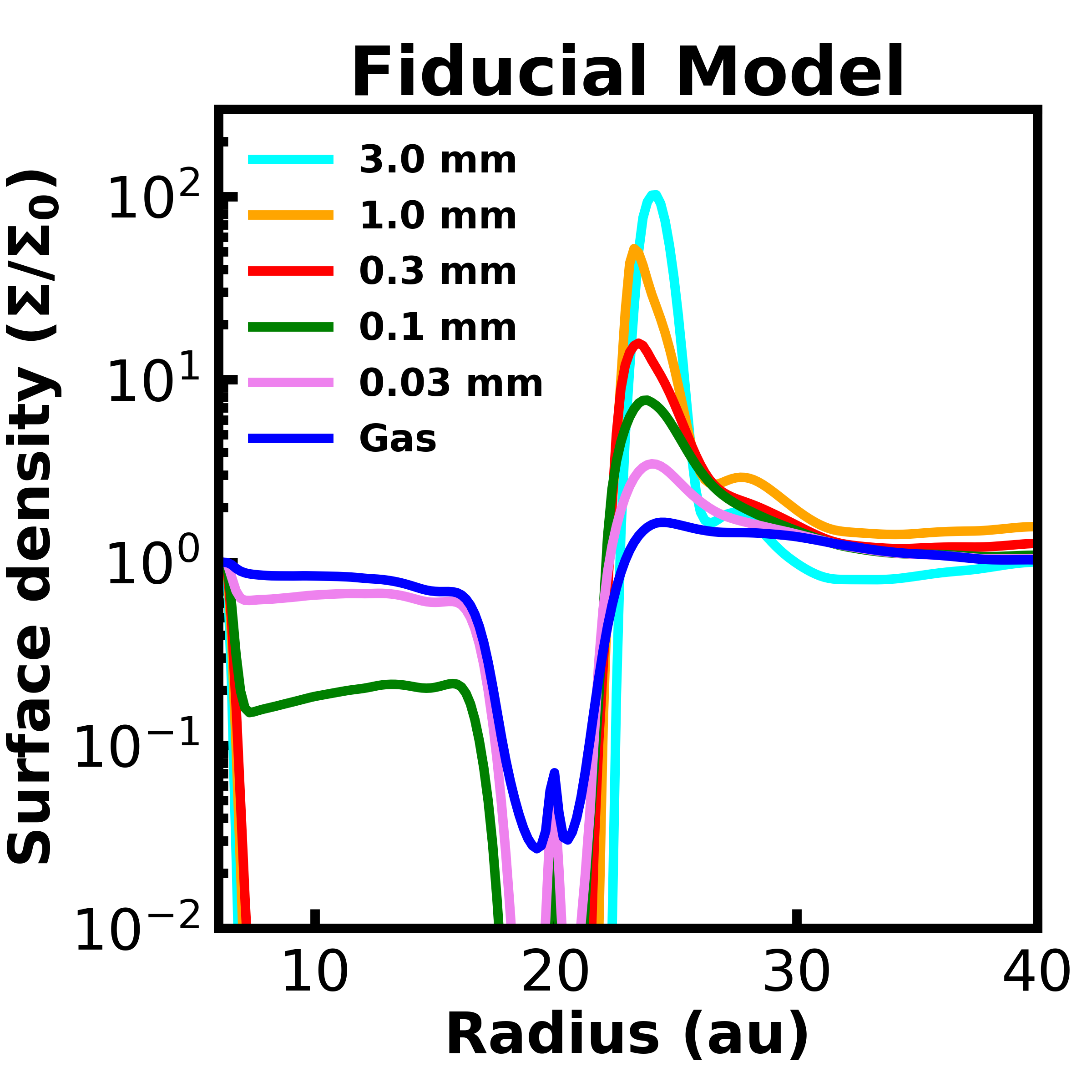}
\caption{The averaged radial profiles for each fluid surface density in fiducial model with a 53 $M_{\oplus}$ planet at 20 au from the central star.}
\label{fig:1d density}
\end{figure}

The MHD disk wind transfers angular momentum at the disk's mid-plane in a manner akin to inward radial drift, \cite{WU_Chen_jiang_2023} believed that during the process of angular momentum transfer, the wind pushes the fluid inward as much as possible. Due to planet-disk interactions, the fluid then accumulates just behind the planet, creating a vortensity gradient (as shown in Fig.\ref{fig:1d density}, see the orange (cyan) line for 1 (3) mm dust). According to RWI, this process makes it easier for vortices to form on the ring compared to a viscosity-dominated disk, manifesting in observations as asymmetric structures with high signal (as shown in the pure wind case in Fig.\ref{fig:reference case}). This is also why in the 7mm JVLA observations, the asymmetric structures appear more significant compared to the observations in ALMA Band 6.

It is important to note that, unlike the traditional vortices which lack bright dust rings (i.e., the pure-wind case in Fig.\ref{fig:reference case}), our fiducial model still retains complete dust ring structures. Upon incorporating appropriate viscosity, the interaction between wind and viscosity tends to break down large clumps into smaller, numerous clumps distributed across the ring. This accounts for the source of asymmetric structures we observe in Fig.\ref{fig:best-fit}. Additionally, appropriate viscosity and a moderate planet mass generally maintain structural symmetry and stability in the dust rings and gaps. This is something that massive-planet models are unable to achieve \citep{ChenKan2024}.

Similar vortex structures also seen in 3D MHD simulations \citep{Aoyama-Bai-2023,Hu2024}, which suggest that the tendency for MHD disk wind to facilitate the formation of vortices/asymmetric structures is not merely an artifact of our wind prescription or the 2D geometric structure used in our simulations.

\subsection{Caveats and Outlook}\label{subsec:caveats}
Unlike real 3D MHD simulations, our simplified 2D MHD disk wind prescription allows for longer run times and a wider range of parameter exploration, and the dust dynamics can be considered in simulations. However, this also means our hydrodynamic simulations are limited by the 2D geometry, with the MHD wind only added as a torque term driving radial gas inflow. This approximation cannot capture feedback effects on the initial magnetic field structure or wind profiles during gap opening.

A recent study conducted a detailed examination using three-dimensional magnetohydrodynamic simulations of star-disk interactions \citep{Aoyama-Bai-2023}. They explored how MHD winds transport gas through planetary gaps and examined the migration torques exerted on the embedded planets by the disk. The conclusion was that due to magnetic flux concentration near the planet, the wind torque within the gap is increased by 3-5 times. This implies that $\alpha_{\rm dw}$ should be stronger in the gap region, naturally resulting in stronger turbulence. However, while these 3D MHD simulations are complex and capture additional physical features, they are computationally expensive and thus cannot run for the long timescales related to ALMA disk ages. They also cannot incorporate dust dynamics to generate simulated observational images for comparison with ALMA observations. This means that in the future, we should look to optimize our 2D MHD disk wind prescription based on the latest 3D simulation results for more refined outcomes.

Additionally, despite many observations of DM Tau, some key parameters like disk mass and scale height remain unclear, yet these parameters are crucial for observational structures. To better model and analyze the evolution of DM Tau theoretically, more critical parameters from observations are needed for better constraints.

It's important to note that our simulations do not cover the inner disk within 4 au of DM Tau, \cite{Hashimoto2021} have found substructures and the potential presence of a type I planet. If there is indeed high turbulence in DM Tau, it's necessary to consider possible planet migration \citep{McNally2020,Kimmig-2020,Ziampras2024,Turpin2024} and strong magnetic field problems, especially in the inner region. \cite{Wu_2024_chaotic} indicates that high turbulence can cause chaotic migration of type I planets, and studies show that planet migration can also affect the formation of substructures \citep{Wu_2023_migration}. Moreover, under appropriate magnetic field conditions, the dust dynamics in the inner region might differ from the usual scenario \citep{Krapp2018,Wu_2024_BDHI}, which future theoretical research needs to consider.

Furthermore, recent $[$O I$]$ line spectral mapping observations have identified evidence of wind in TW Hya \citep{Fang2023_TWHya_wind}, PDS 70 \citep{Campbell-White2023_PDS70} and T Cha \citep{Bajaj2024_TCha}. Though there is still debate over whether it's MHD disk wind or photoevaporative wind \citep{Lin_Wang_2024,Sellek2024}. However, our study suggests DM Tau is undoubtedly a source suitable for future related observations.

\section{Conclusions}\label{sec: conclusions}
In this work, we present the results of multi-fluid hydrodynamic simulations that incorporate the 2D prescription of MHD disk wind. Through planet-disk interaction and wind-viscosity interaction, we successfully fit some observational characteristics of DM Tau in different wavelength (ALMA Band 6 and JVLA Q Band), thereby investigating the source of high accretion rate in this system. Our main findings are as follows:

\begin{itemize}
    \item We find that the high accretion rate in DM Tau could arise from the combined action of MHD disk wind and viscosity, where MHD disk wind should account for about $97\%$ of the contribution, with the remaining approximately $3\%$ coming from viscosity. The high-contrast clumps that produce asymmetric structures in the bright dust ring around 21 au in DM Tau, as well as gap opening, could be mainly result from the interaction between the wind and a 53 $M_{\oplus}$ planet, affecting the distribution of dust. While the maintenance of the ring and gap morphology comes from appropriate viscosity. Based on suggestions from MHD simulations \citep{Bai2017}, this may imply that the potential Hall effect in DM Tau cannot be ignored.

    \item Our study indicates that a planet fitting the material mass of the DM Tau system can indeed exist within the 4-21 au gap in DM Tau, and the formation of a gap consistent with observational structures does not require the presence of a gas giant planet as large as 1 $M_J$ or even 10 $M_J$.
    
    \item Our research also supports recent conclusions from sub-cm wavelength observations of DM Tau \citep{Baobab_2024}, namely that at wavelengths $\textless$ 1.3 mm, the flux density is dominated by optically thick dust thermal emission, while at sub-mm and even longer wavelengths, the flux density should difference.
     
\end{itemize}

Although \cite{Flaherty2020} calculated an alpha value of $\alpha_{\rm v}\sim 0.078$ by detecting nonthermal gas motions in DM Tau, indicating that DM Tau is a highly turbulent system, our results do not actually conflict with this. This is because the turbulence level measured by \cite{Flaherty2020} comes from regions beyond 100 au from DM Tau and above the midplane ($\sim 2h$), whereas our focus is on the midplane and the region within $\sim$ 20 au. Moreover, MHD simulations also suggest that the turbulence level at the midplane should be lower than that in higher regions \citep{Simon2015}.

Overall, as a system with such a high accretion rate and varied substructures, DM Tau warrants further research effort in the future, especially with the help of upcoming observational facilities like SKA and ngVLA in longer wavelength \citep{Ricci2018,Ilee2020,Ricci2021,Wu_2024_ska_ngvla}. These will complement the existing capabilities of ALMA and JVLA in terms of wavelength or resolution, thus better constraining key parameters of the disk (e.g. disk mass and temperature). Meanwhile, MHD disk wind, as an important addition to existing accretion and planet formation theories, deserves to be applied across a broader range of systems.

\section*{Aknowledgement}
Y.W. thanks the anonymous referee for highly constructive suggestions, which significantly enhanced the quality of the manuscript. Y.W. thanks Jun Hashimoto and Hauyu Baobab Liu for their discussions and share observational data. Y.W. also thanks Xue-Ning Bai, Kan Chen, Yi-Xian Chen and Wenrui Xu for very useful suggestions.

This research used DiRAC Data Intensive service at Leicester, operated by the University of Leicester IT Services, which forms part of the STFC DiRAC HPC Facility (\href{www.dirac.ac.uk}{www.dirac.ac.uk}).

\section*{Data availability}
The data obtained in our simulations can be made available on reasonable request to the corresponding author. 

\appendix
\section{Radiative Transfer Calculations}\label{appendix: RT}
During the process of performing radiative transfer calculations on the results of hydrodynamic simulations using \texttt{RADMC-3D}, we first model the vertical expansion of the 2D dust density map into 3D by adopting follow Gaussian distribution for the dust densities along the vertical axis with 40 grids:
\begin{equation}
  \rho_{\rm d}(r,z) = \Sigma(r)\exp(-z^2/2h_{\rm d}^2)) / \sqrt{2\pi}.
  \label{eq:rho_d}
\end{equation}

To process the dust scale height, $h_d$, a common method is using the formula 
\begin{equation}
  h_d \approx \sqrt{D_{\rm g}/(\Omega {\rm St})},
\end{equation} 
as detailed by \cite{Dubrulle1995,Youdin-Lithwick}, here $D_{\rm g}$ is the vertical turbulent diffusion coefficient at the midplane.

As shown in Eqs. \ref{HD-dens-gas} - \ref{HD-vel-dust}, since our code is multi-fluid, the wind torque does not directly interact with the dust in the Navier-Stokes equations. That means our wind torque does it affect the diffusion evolution of the dust itself in simulations. Therefore, we have ignored all contributions of MHD winds to dust diffusion.

The coefficient $D_{\rm g}$, is often equated to the viscosity $\nu = \alpha_{\rm v}c_s h_{\rm g}$ under the assumption of isotropic turbulence \citep{Birnstiel2016,Baruteau21}. Yet, findings from MHD simulations introduce a layer of complexity \citep{Zhu-2015,Bai2016}, highlighting the potential inaccuracies in this assumption. The presence of wind can elevate dust to higher altitudes, however, this typically applies to micron-sized dust with very small Stokes numbers, which have $h_d$ intrinsically reaching the wind base at $\sim h$ \citep{Armitage2013}, rather than to millimeter-sized dust.

Our analysis, particularly focused on disks observed face-on, has shown that variations in the vertical diffusion coefficient minimally impact the observed results. Thus, for simplicity and to concentrate on analyzing data from relatively face-on perspectives, we adopt a uniform dust scale height $h_{\rm d} = 0.1~h_{\rm g}$ across all dust components, aligning with the approach of \cite{Miranda2017}. This adjustment equates to a diffusivity coefficient $D_{\rm g} = 10^{-4}h_g^2\Omega$ for pebbles with a Stokes number of 0.01. Simulations incorporating both ideal and non-ideal MHD effects \citep{Bai2014,BaehrZhu2021,Zhou2022} suggesting possible variations in dust diffusivity along vertical and radial dimensions.

However, some works indicate that the vertical setting of particles can affect several observables, such as the total millimeter flux and shape of the substructures, even if the disk inclination is low \citep[e.g.,][]{Pinilla_2021}. This implies that we should explore this aspect further in future research.

We apply the DSHARP opacities version without ice, it assume our dust species comprised of $41.1\%$ astro-silicates, 49.6\% CHON organics and $9.3\%$ FeS, with internal density $\rho_{\rm d}=2.11~g/cm^{3}$ consistent with our simulations. 

\bibliography{DMTau}{}

\begin{thebibliography}{}
\expandafter\ifx\csname natexlab\endcsname\relax\def\natexlab#1{#1}\fi
\providecommand{\url}[1]{\href{#1}{#1}}
\providecommand{\dodoi}[1]{doi:~\href{http://doi.org/#1}{\nolinkurl{#1}}}
\providecommand{\doeprint}[1]{\href{http://ascl.net/#1}{\nolinkurl{http://ascl.net/#1}}}
\providecommand{\doarXiv}[1]{\href{https://arxiv.org/abs/#1}{\nolinkurl{https://arxiv.org/abs/#1}}}

\bibitem[{{ALMA Partnership} {et~al.}(2015){ALMA Partnership}, {Brogan}, {P{\'e}rez}, {Hunter}, {Dent}, {Hales}, {Hills}, {Corder}, {Fomalont}, {Vlahakis}, {Asaki}, {Barkats}, {Hirota}, {Hodge}, {Impellizzeri}, {Kneissl}, {Liuzzo}, {Lucas}, {Marcelino}, {Matsushita}, {Nakanishi}, {Phillips}, {Richards}, {Toledo}, {Aladro}, {Broguiere}, {Cortes}, {Cortes}, {Espada}, {Galarza}, {Garcia-Appadoo}, {Guzman-Ramirez}, {Humphreys}, {Jung}, {Kameno}, {Laing}, {Leon}, {Marconi}, {Mignano}, {Nikolic}, {Nyman}, {Radiszcz}, {Remijan}, {Rod{\'o}n}, {Sawada}, {Takahashi}, {Tilanus}, {Vila Vilaro}, {Watson}, {Wiklind}, {Akiyama}, {Chapillon}, {de Gregorio-Monsalvo}, {Di Francesco}, {Gueth}, {Kawamura}, {Lee}, {Nguyen Luong}, {Mangum}, {Pietu}, {Sanhueza}, {Saigo}, {Takakuwa}, {Ubach}, {van Kempen}, {Wootten}, {Castro-Carrizo}, {Francke}, {Gallardo}, {Garcia}, {Gonzalez}, {Hill}, {Kaminski}, {Kurono}, {Liu}, {Lopez}, {Morales}, {Plarre}, {Schieven}, {Testi}, {Videla}, {Villard}, {Andreani}, {Hibbard}, \&
  {Tatematsu}}]{ALMA2015}
{ALMA Partnership}, {Brogan}, C.~L., {P{\'e}rez}, L.~M., {et~al.} 2015, \apjl, 808, L3, \dodoi{10.1088/2041-8205/808/1/L3}

\bibitem[{{Andrews}(2020)}]{Andrews2020}
{Andrews}, S.~M. 2020, \araa, 58, 483, \dodoi{10.1146/annurev-astro-031220-010302}

\bibitem[{{Andrews} {et~al.}(2018){Andrews}, {Huang}, {P{\'e}rez}, {Isella}, {Dullemond}, {Kurtovic}, {Guzm{\'a}n}, {Carpenter}, {Wilner}, {Zhang}, {Zhu}, {Birnstiel}, {Bai}, {Benisty}, {Hughes}, {{\"O}berg}, \& {Ricci}}]{andrews2018}
{Andrews}, S.~M., {Huang}, J., {P{\'e}rez}, L.~M., {et~al.} 2018, \apjl, 869, L41, \dodoi{10.3847/2041-8213/aaf741}

\bibitem[{{Aoyama} \& {Bai}(2023)}]{Aoyama-Bai-2023}
{Aoyama}, Y., \& {Bai}, X.-N. 2023, \apj, 946, 5, \dodoi{10.3847/1538-4357/acb81f}

\bibitem[{{Armitage} {et~al.}(2013){Armitage}, {Simon}, \& {Martin}}]{Armitage2013}
{Armitage}, P.~J., {Simon}, J.~B., \& {Martin}, R.~G. 2013, \apjl, 778, L14, \dodoi{10.1088/2041-8205/778/1/L14}

\bibitem[{{Baehr} \& {Zhu}(2021)}]{BaehrZhu2021}
{Baehr}, H., \& {Zhu}, Z. 2021, \apj, 909, 136, \dodoi{10.3847/1538-4357/abddb4}

\bibitem[{{Bai}(2011)}]{Bai2011}
{Bai}, X.-N. 2011, \apj, 739, 50, \dodoi{10.1088/0004-637X/739/1/50}

\bibitem[{{Bai}(2013)}]{Bai2013}
---. 2013, \apj, 772, 96, \dodoi{10.1088/0004-637X/772/2/96}

\bibitem[{{Bai}(2015)}]{Bai2015}
---. 2015, \apj, 798, 84, \dodoi{10.1088/0004-637X/798/2/84}

\bibitem[{{Bai} \& {Stone}(2013)}]{2013BaiStone}
{Bai}, X.-N., \& {Stone}, J.~M. 2013, \apj, 769, 76, \dodoi{10.1088/0004-637X/769/1/76}

\bibitem[{{Bai} \& {Stone}(2014)}]{Bai2014}
---. 2014, \apj, 796, 31, \dodoi{10.1088/0004-637X/796/1/31}

\bibitem[{{Bai} \& {Stone}(2017)}]{Bai2017}
---. 2017, \apj, 836, 46, \dodoi{10.3847/1538-4357/836/1/46}

\bibitem[{{Bai} {et~al.}(2016){Bai}, {Ye}, {Goodman}, \& {Yuan}}]{Bai2016}
{Bai}, X.-N., {Ye}, J., {Goodman}, J., \& {Yuan}, F. 2016, \apj, 818, 152, \dodoi{10.3847/0004-637X/818/2/152}

\bibitem[{{Bajaj} {et~al.}(2024){Bajaj}, {Pascucci}, {Gorti}, {Alexander}, {Sellek}, {Morrison}, {Gaspar}, {Clarke}, {Xie}, {Ballabio}, \& {Deng}}]{Bajaj2024_TCha}
{Bajaj}, N.~S., {Pascucci}, I., {Gorti}, U., {et~al.} 2024, \aj, 167, 127, \dodoi{10.3847/1538-3881/ad22e1}

\bibitem[{{Baruteau} {et~al.}(2021){Baruteau}, {Wafflard-Fernandez}, {Le Gal}, {Debras}, {Carmona}, {Fuente}, \& {Rivi{\`e}re-Marichalar}}]{Baruteau21}
{Baruteau}, C., {Wafflard-Fernandez}, G., {Le Gal}, R., {et~al.} 2021, \mnras, 505, 359, \dodoi{10.1093/mnras/stab1045}

\bibitem[{{Baruteau} {et~al.}(2019){Baruteau}, {Barraza}, {P{\'e}rez}, {Casassus}, {Dong}, {Lyra}, {Marino}, {Christiaens}, {Zhu}, {Carmona}, {Debras}, \& {Alarcon}}]{fargo2radmc3d-dust}
{Baruteau}, C., {Barraza}, M., {P{\'e}rez}, S., {et~al.} 2019, \mnras, 486, 304, \dodoi{10.1093/mnras/stz802}

\bibitem[{{Ben{\'\i}tez-Llambay} {et~al.}(2019){Ben{\'\i}tez-Llambay}, {Krapp}, \& {Pessah}}]{FARGO3D-multifluid}
{Ben{\'\i}tez-Llambay}, P., {Krapp}, L., \& {Pessah}, M.~E. 2019, \apjs, 241, 25, \dodoi{10.3847/1538-4365/ab0a0e}

\bibitem[{{Ben{\'\i}tez-Llambay} \& {Masset}(2016)}]{FARGO3D}
{Ben{\'\i}tez-Llambay}, P., \& {Masset}, F.~S. 2016, \apjs, 223, 11, \dodoi{10.3847/0067-0049/223/1/11}

\bibitem[{{Birnstiel} {et~al.}(2016){Birnstiel}, {Fang}, \& {Johansen}}]{Birnstiel2016}
{Birnstiel}, T., {Fang}, M., \& {Johansen}, A. 2016, \ssr, 205, 41, \dodoi{10.1007/s11214-016-0256-1}

\bibitem[{{Birnstiel} {et~al.}(2018){Birnstiel}, {Dullemond}, {Zhu}, {Andrews}, {Bai}, {Wilner}, {Carpenter}, {Huang}, {Isella}, {Benisty}, {P{\'e}rez}, \& {Zhang}}]{DSHARP-V}
{Birnstiel}, T., {Dullemond}, C.~P., {Zhu}, Z., {et~al.} 2018, \apjl, 869, L45, \dodoi{10.3847/2041-8213/aaf743}

\bibitem[{{Campbell-White} {et~al.}(2023){Campbell-White}, {Manara}, {Benisty}, {Natta}, {Claes}, {Frasca}, {Bae}, {Facchini}, {Isella}, {P{\'e}rez}, {Pinilla}, {Sicilia-Aguilar}, \& {Teague}}]{Campbell-White2023_PDS70}
{Campbell-White}, J., {Manara}, C.~F., {Benisty}, M., {et~al.} 2023, \apj, 956, 25, \dodoi{10.3847/1538-4357/acf0c0}

\bibitem[{{Chen} {et~al.}(2024){Chen}, {Kama}, {Pinilla}, \& {Keyte}}]{ChenKan2024}
{Chen}, K., {Kama}, M., {Pinilla}, P., \& {Keyte}, L. 2024, \mnras, 527, 2049, \dodoi{10.1093/mnras/stad3247}

\bibitem[{{Chen} {et~al.}(2021){Chen}, {Wang}, {Li}, {Baruteau}, \& {Lin}}]{Chen2021}
{Chen}, Y.-X., {Wang}, Z., {Li}, Y.-P., {Baruteau}, C., \& {Lin}, D. N.~C. 2021, \apj, 922, 184, \dodoi{10.3847/1538-4357/ac23d7}

\bibitem[{{Chung} {et~al.}(2024){Chung}, {Andrews}, {Gurwell}, {Wright}, {Long}, {Xu}, \& {Liu}}]{Chung2024}
{Chung}, C.-Y., {Andrews}, S.~M., {Gurwell}, M.~A., {et~al.} 2024, arXiv e-prints, arXiv:2405.19867, \dodoi{10.48550/arXiv.2405.19867}

\bibitem[{{Crida} {et~al.}(2006){Crida}, {Morbidelli}, \& {Masset}}]{Crida2006}
{Crida}, A., {Morbidelli}, A., \& {Masset}, F. 2006, \icarus, 181, 587, \dodoi{10.1016/j.icarus.2005.10.007}

\bibitem[{{Cui} \& {Bai}(2021)}]{Cui2021}
{Cui}, C., \& {Bai}, X.-N. 2021, \mnras, 507, 1106, \dodoi{10.1093/mnras/stab2220}

\bibitem[{{Cui} \& {Bai}(2022)}]{Cui2022}
---. 2022, \mnras, 516, 4660, \dodoi{10.1093/mnras/stac2580}

\bibitem[{{Delage} {et~al.}(2022){Delage}, {Okuzumi}, {Flock}, {Pinilla}, \& {Dzyurkevich}}]{Delage_2022_MRI_Viscosity}
{Delage}, T.~N., {Okuzumi}, S., {Flock}, M., {Pinilla}, P., \& {Dzyurkevich}, N. 2022, \aap, 658, A97, \dodoi{10.1051/0004-6361/202141689}

\bibitem[{{Doi} \& {Kataoka}(2021)}]{DoiKataoka2021}
{Doi}, K., \& {Kataoka}, A. 2021, \apj, 912, 164, \dodoi{10.3847/1538-4357/abe5a6}

\bibitem[{{Dominik} {et~al.}(2021){Dominik}, {Min}, \& {Tazaki}}]{optool-2021}
{Dominik}, C., {Min}, M., \& {Tazaki}, R. 2021, {OpTool: Command-line driven tool for creating complex dust opacities}, Astrophysics Source Code Library, record ascl:2104.010.
\newblock \doeprint{2104.010}

\bibitem[{{Dong} {et~al.}(2017){Dong}, {Li}, {Chiang}, \& {Li}}]{Dong_2017_super-earth}
{Dong}, R., {Li}, S., {Chiang}, E., \& {Li}, H. 2017, \apj, 843, 127, \dodoi{10.3847/1538-4357/aa72f2}

\bibitem[{{Dong} {et~al.}(2018){Dong}, {Li}, {Chiang}, \& {Li}}]{Dong_2018_super-earth}
---. 2018, \apj, 866, 110, \dodoi{10.3847/1538-4357/aadadd}

\bibitem[{{Dubrulle} {et~al.}(1995){Dubrulle}, {Morfill}, \& {Sterzik}}]{Dubrulle1995}
{Dubrulle}, B., {Morfill}, G., \& {Sterzik}, M. 1995, \icarus, 114, 237, \dodoi{10.1006/icar.1995.1058}

\bibitem[{{Dullemond} {et~al.}(2012){Dullemond}, {Juhasz}, {Pohl}, {Sereshti}, {Shetty}, {Peters}, {Commercon}, \& {Flock}}]{RADMC-3D}
{Dullemond}, C.~P., {Juhasz}, A., {Pohl}, A., {et~al.} 2012, {RADMC-3D: A multi-purpose radiative transfer tool}.
\newblock \doeprint{1202.015}

\bibitem[{{Dullemond} {et~al.}(2018){Dullemond}, {Birnstiel}, {Huang}, {Kurtovic}, {Andrews}, {Guzm{\'a}n}, {P{\'e}rez}, {Isella}, {Zhu}, {Benisty}, {Wilner}, {Bai}, {Carpenter}, {Zhang}, \& {Ricci}}]{Dullemond2018}
{Dullemond}, C.~P., {Birnstiel}, T., {Huang}, J., {et~al.} 2018, \apjl, 869, L46, \dodoi{10.3847/2041-8213/aaf742}

\bibitem[{{Elbakyan} {et~al.}(2022){Elbakyan}, {Wu}, {Nayakshin}, \& {Rosotti}}]{MHD-wind-Elbakyan}
{Elbakyan}, V., {Wu}, Y., {Nayakshin}, S., \& {Rosotti}, G. 2022, \mnras, 515, 3113, \dodoi{10.1093/mnras/stac1774}

\bibitem[{{Fang} {et~al.}(2023){Fang}, {Wang}, {Herczeg}, {Hashimoto}, {Xu}, {Nemer}, {Pascucci}, {Haffert}, \& {Aoyama}}]{Fang2023_TWHya_wind}
{Fang}, M., {Wang}, L., {Herczeg}, G.~J., {et~al.} 2023, Nature Astronomy, 7, 905, \dodoi{10.1038/s41550-023-02004-x}

\bibitem[{{Flaherty} {et~al.}(2020){Flaherty}, {Hughes}, {Simon}, {Qi}, {Bai}, {Bulatek}, {Andrews}, {Wilner}, \& {K{\'o}sp{\'a}l}}]{Flaherty2020}
{Flaherty}, K., {Hughes}, A.~M., {Simon}, J.~B., {et~al.} 2020, \apj, 895, 109, \dodoi{10.3847/1538-4357/ab8cc5}

\bibitem[{{Flaherty} {et~al.}(2017){Flaherty}, {Hughes}, {Rose}, {Simon}, {Qi}, {Andrews}, {K{\'o}sp{\'a}l}, {Wilner}, {Chiang}, {Armitage}, \& {Bai}}]{Flaherty2017}
{Flaherty}, K.~M., {Hughes}, A.~M., {Rose}, S.~C., {et~al.} 2017, \apj, 843, 150, \dodoi{10.3847/1538-4357/aa79f9}

\bibitem[{{Francis} \& {van der Marel}(2020)}]{Francis_Nienke_2020}
{Francis}, L., \& {van der Marel}, N. 2020, \apj, 892, 111, \dodoi{10.3847/1538-4357/ab7b63}

\bibitem[{{Francis} {et~al.}(2022){Francis}, {Marel}, {Johnstone}, {Akiyama}, {Bruderer}, {Dong}, {Hashimoto}, {Liu}, {Muto}, \& {Yang}}]{Francis2022}
{Francis}, L., {Marel}, N. v.~d., {Johnstone}, D., {et~al.} 2022, \aj, 164, 105, \dodoi{10.3847/1538-3881/ac7ffb}

\bibitem[{{Gaia Collaboration} {et~al.}(2021){Gaia Collaboration}, {Brown}, {Vallenari}, {Prusti}, {de Bruijne}, {Babusiaux}, {Biermann}, {Creevey}, {Evans}, {Eyer}, {Hutton}, {Jansen}, {Jordi}, {Klioner}, {Lammers}, {Lindegren}, {Luri}, {Mignard}, {Panem}, {Pourbaix}, {Randich}, {Sartoretti}, {Soubiran}, {Walton}, {Arenou}, {Bailer-Jones}, {Bastian}, {Cropper}, {Drimmel}, {Katz}, {Lattanzi}, {van Leeuwen}, {Bakker}, {Cacciari}, {Casta{\~n}eda}, {De Angeli}, {Ducourant}, {Fabricius}, {Fouesneau}, {Fr{\'e}mat}, {Guerra}, {Guerrier}, {Guiraud}, {Jean-Antoine Piccolo}, {Masana}, {Messineo}, {Mowlavi}, {Nicolas}, {Nienartowicz}, {Pailler}, {Panuzzo}, {Riclet}, {Roux}, {Seabroke}, {Sordo}, {Tanga}, {Th{\'e}venin}, {Gracia-Abril}, {Portell}, {Teyssier}, {Altmann}, {Andrae}, {Bellas-Velidis}, {Benson}, {Berthier}, {Blomme}, {Brugaletta}, {Burgess}, {Busso}, {Carry}, {Cellino}, {Cheek}, {Clementini}, {Damerdji}, {Davidson}, {Delchambre}, {Dell'Oro}, {Fern{\'a}ndez-Hern{\'a}ndez}, {Galluccio}, {Garc{\'\i}a-Lario},
  {Garcia-Reinaldos}, {Gonz{\'a}lez-N{\'u}{\~n}ez}, {Gosset}, {Haigron}, {Halbwachs}, {Hambly}, {Harrison}, {Hatzidimitriou}, {Heiter}, {Hern{\'a}ndez}, {Hestroffer}, {Hodgkin}, {Holl}, {Jan{\ss}en}, {Jevardat de Fombelle}, {Jordan}, {Krone-Martins}, {Lanzafame}, {L{\"o}ffler}, {Lorca}, {Manteiga}, {Marchal}, {Marrese}, {Moitinho}, {Mora}, {Muinonen}, {Osborne}, {Pancino}, {Pauwels}, {Petit}, {Recio-Blanco}, {Richards}, {Riello}, {Rimoldini}, {Robin}, {Roegiers}, {Rybizki}, {Sarro}, {Siopis}, {Smith}, {Sozzetti}, {Ulla}, {Utrilla}, {van Leeuwen}, {van Reeven}, {Abbas}, {Abreu Aramburu}, {Accart}, {Aerts}, {Aguado}, {Ajaj}, {Altavilla}, {{\'A}lvarez}, {{\'A}lvarez Cid-Fuentes}, {Alves}, {Anderson}, {Anglada Varela}, {Antoja}, {Audard}, {Baines}, {Baker}, {Balaguer-N{\'u}{\~n}ez}, {Balbinot}, {Balog}, {Barache}, {Barbato}, {Barros}, {Barstow}, {Bartolom{\'e}}, {Bassilana}, {Bauchet}, {Baudesson-Stella}, {Becciani}, {Bellazzini}, {Bernet}, {Bertone}, {Bianchi}, {Blanco-Cuaresma}, {Boch}, {Bombrun}, {Bossini},
  {Bouquillon}, {Bragaglia}, {Bramante}, {Breedt}, {Bressan}, {Brouillet}, {Bucciarelli}, {Burlacu}, {Busonero}, {Butkevich}, {Buzzi}, {Caffau}, {Cancelliere}, {C{\'a}novas}, {Cantat-Gaudin}, {Carballo}, {Carlucci}, {Carnerero}, {Carrasco}, {Casamiquela}, {Castellani}, {Castro-Ginard}, {Castro Sampol}, {Chaoul}, {Charlot}, {Chemin}, {Chiavassa}, {Cioni}, {Comoretto}, {Cooper}, {Cornez}, {Cowell}, {Crifo}, {Crosta}, {Crowley}, {Dafonte}, {Dapergolas}, {David}, {David}, {de Laverny}, {De Luise}, {De March}, {De Ridder}, {de Souza}, {de Teodoro}, {de Torres}, {del Peloso}, {del Pozo}, {Delbo}, {Delgado}, {Delgado}, {Delisle}, {Di Matteo}, {Diakite}, {Diener}, {Distefano}, {Dolding}, {Eappachen}, {Edvardsson}, {Enke}, {Esquej}, {Fabre}, {Fabrizio}, {Faigler}, {Fedorets}, {Fernique}, {Fienga}, {Figueras}, {Fouron}, {Fragkoudi}, {Fraile}, {Franke}, {Gai}, {Garabato}, {Garcia-Gutierrez}, {Garc{\'\i}a-Torres}, {Garofalo}, {Gavras}, {Gerlach}, {Geyer}, {Giacobbe}, {Gilmore}, {Girona}, {Giuffrida}, {Gomel}, {Gomez},
  {Gonzalez-Santamaria}, {Gonz{\'a}lez-Vidal}, {Granvik}, {Guti{\'e}rrez-S{\'a}nchez}, {Guy}, {Hauser}, {Haywood}, {Helmi}, {Hidalgo}, {Hilger}, {H{\l}adczuk}, {Hobbs}, {Holland}, {Huckle}, {Jasniewicz}, {Jonker}, {Juaristi Campillo}, {Julbe}, {Karbevska}, {Kervella}, {Khanna}, {Kochoska}, {Kontizas}, {Kordopatis}, {Korn}, {Kostrzewa-Rutkowska}, {Kruszy{\'n}ska}, {Lambert}, {Lanza}, {Lasne}, {Le Campion}, {Le Fustec}, {Lebreton}, {Lebzelter}, {Leccia}, {Leclerc}, {Lecoeur-Taibi}, {Liao}, {Licata}, {Lindstr{\o}m}, {Lister}, {Livanou}, {Lobel}, {Madrero Pardo}, {Managau}, {Mann}, {Marchant}, {Marconi}, {Marcos Santos}, {Marinoni}, {Marocco}, {Marshall}, {Martin Polo}, {Mart{\'\i}n-Fleitas}, {Masip}, {Massari}, {Mastrobuono-Battisti}, {Mazeh}, {McMillan}, {Messina}, {Michalik}, {Millar}, {Mints}, {Molina}, {Molinaro}, {Moln{\'a}r}, {Montegriffo}, {Mor}, {Morbidelli}, {Morel}, {Morris}, {Mulone}, {Munoz}, {Muraveva}, {Murphy}, {Musella}, {Noval}, {Ord{\'e}novic}, {Orr{\`u}}, {Osinde}, {Pagani}, {Pagano},
  {Palaversa}, {Palicio}, {Panahi}, {Pawlak}, {Pe{\~n}alosa Esteller}, {Penttil{\"a}}, {Piersimoni}, {Pineau}, {Plachy}, {Plum}, {Poggio}, {Poretti}, {Poujoulet}, {Pr{\v{s}}a}, {Pulone}, {Racero}, {Ragaini}, {Rainer}, {Raiteri}, {Rambaux}, {Ramos}, {Ramos-Lerate}, {Re Fiorentin}, {Regibo}, {Reyl{\'e}}, {Ripepi}, {Riva}, {Rixon}, {Robichon}, {Robin}, {Roelens}, {Rohrbasser}, {Romero-G{\'o}mez}, {Rowell}, {Royer}, {Rybicki}, {Sadowski}, {Sagrist{\`a} Sell{\'e}s}, {Sahlmann}, {Salgado}, {Salguero}, {Samaras}, {Sanchez Gimenez}, {Sanna}, {Santove{\~n}a}, {Sarasso}, {Schultheis}, {Sciacca}, {Segol}, {Segovia}, {S{\'e}gransan}, {Semeux}, {Shahaf}, {Siddiqui}, {Siebert}, {Siltala}, {Slezak}, {Smart}, {Solano}, {Solitro}, {Souami}, {Souchay}, {Spagna}, {Spoto}, {Steele}, {Steidelm{\"u}ller}, {Stephenson}, {S{\"u}veges}, {Szabados}, {Szegedi-Elek}, {Taris}, {Tauran}, {Taylor}, {Teixeira}, {Thuillot}, {Tonello}, {Torra}, {Torra}, {Turon}, {Unger}, {Vaillant}, {van Dillen}, {Vanel}, {Vecchiato}, {Viala}, {Vicente},
  {Voutsinas}, {Weiler}, {Wevers}, {Wyrzykowski}, {Yoldas}, {Yvard}, {Zhao}, {Zorec}, {Zucker}, {Zurbach}, \& {Zwitter}}]{Gaia2021}
{Gaia Collaboration}, {Brown}, A.~G.~A., {Vallenari}, A., {et~al.} 2021, \aap, 649, A1, \dodoi{10.1051/0004-6361/202039657}

\bibitem[{{Gaia Collaboration} {et~al.}(2023){Gaia Collaboration}, {Vallenari}, {Brown}, {Prusti}, {de Bruijne}, {Arenou}, {Babusiaux}, {Biermann}, {Creevey}, {Ducourant}, {Evans}, {Eyer}, {Guerra}, {Hutton}, {Jordi}, {Klioner}, {Lammers}, {Lindegren}, {Luri}, {Mignard}, {Panem}, {Pourbaix}, {Randich}, {Sartoretti}, {Soubiran}, {Tanga}, {Walton}, {Bailer-Jones}, {Bastian}, {Drimmel}, {Jansen}, {Katz}, {Lattanzi}, {van Leeuwen}, {Bakker}, {Cacciari}, {Casta{\~n}eda}, {De Angeli}, {Fabricius}, {Fouesneau}, {Fr{\'e}mat}, {Galluccio}, {Guerrier}, {Heiter}, {Masana}, {Messineo}, {Mowlavi}, {Nicolas}, {Nienartowicz}, {Pailler}, {Panuzzo}, {Riclet}, {Roux}, {Seabroke}, {Sordo}, {Th{\'e}venin}, {Gracia-Abril}, {Portell}, {Teyssier}, {Altmann}, {Andrae}, {Audard}, {Bellas-Velidis}, {Benson}, {Berthier}, {Blomme}, {Burgess}, {Busonero}, {Busso}, {C{\'a}novas}, {Carry}, {Cellino}, {Cheek}, {Clementini}, {Damerdji}, {Davidson}, {de Teodoro}, {Nu{\~n}ez Campos}, {Delchambre}, {Dell'Oro}, {Esquej},
  {Fern{\'a}ndez-Hern{\'a}ndez}, {Fraile}, {Garabato}, {Garc{\'\i}a-Lario}, {Gosset}, {Haigron}, {Halbwachs}, {Hambly}, {Harrison}, {Hern{\'a}ndez}, {Hestroffer}, {Hodgkin}, {Holl}, {Jan{\ss}en}, {Jevardat de Fombelle}, {Jordan}, {Krone-Martins}, {Lanzafame}, {L{\"o}ffler}, {Marchal}, {Marrese}, {Moitinho}, {Muinonen}, {Osborne}, {Pancino}, {Pauwels}, {Recio-Blanco}, {Reyl{\'e}}, {Riello}, {Rimoldini}, {Roegiers}, {Rybizki}, {Sarro}, {Siopis}, {Smith}, {Sozzetti}, {Utrilla}, {van Leeuwen}, {Abbas}, {{\'A}brah{\'a}m}, {Abreu Aramburu}, {Aerts}, {Aguado}, {Ajaj}, {Aldea-Montero}, {Altavilla}, {{\'A}lvarez}, {Alves}, {Anders}, {Anderson}, {Anglada Varela}, {Antoja}, {Baines}, {Baker}, {Balaguer-N{\'u}{\~n}ez}, {Balbinot}, {Balog}, {Barache}, {Barbato}, {Barros}, {Barstow}, {Bartolom{\'e}}, {Bassilana}, {Bauchet}, {Becciani}, {Bellazzini}, {Berihuete}, {Bernet}, {Bertone}, {Bianchi}, {Binnenfeld}, {Blanco-Cuaresma}, {Blazere}, {Boch}, {Bombrun}, {Bossini}, {Bouquillon}, {Bragaglia}, {Bramante}, {Breedt},
  {Bressan}, {Brouillet}, {Brugaletta}, {Bucciarelli}, {Burlacu}, {Butkevich}, {Buzzi}, {Caffau}, {Cancelliere}, {Cantat-Gaudin}, {Carballo}, {Carlucci}, {Carnerero}, {Carrasco}, {Casamiquela}, {Castellani}, {Castro-Ginard}, {Chaoul}, {Charlot}, {Chemin}, {Chiaramida}, {Chiavassa}, {Chornay}, {Comoretto}, {Contursi}, {Cooper}, {Cornez}, {Cowell}, {Crifo}, {Cropper}, {Crosta}, {Crowley}, {Dafonte}, {Dapergolas}, {David}, {David}, {de Laverny}, {De Luise}, {De March}, {De Ridder}, {de Souza}, {de Torres}, {del Peloso}, {del Pozo}, {Delbo}, {Delgado}, {Delisle}, {Demouchy}, {Dharmawardena}, {Di Matteo}, {Diakite}, {Diener}, {Distefano}, {Dolding}, {Edvardsson}, {Enke}, {Fabre}, {Fabrizio}, {Faigler}, {Fedorets}, {Fernique}, {Fienga}, {Figueras}, {Fournier}, {Fouron}, {Fragkoudi}, {Gai}, {Garcia-Gutierrez}, {Garcia-Reinaldos}, {Garc{\'\i}a-Torres}, {Garofalo}, {Gavel}, {Gavras}, {Gerlach}, {Geyer}, {Giacobbe}, {Gilmore}, {Girona}, {Giuffrida}, {Gomel}, {Gomez}, {Gonz{\'a}lez-N{\'u}{\~n}ez},
  {Gonz{\'a}lez-Santamar{\'\i}a}, {Gonz{\'a}lez-Vidal}, {Granvik}, {Guillout}, {Guiraud}, {Guti{\'e}rrez-S{\'a}nchez}, {Guy}, {Hatzidimitriou}, {Hauser}, {Haywood}, {Helmer}, {Helmi}, {Sarmiento}, {Hidalgo}, {Hilger}, {H{\l}adczuk}, {Hobbs}, {Holland}, {Huckle}, {Jardine}, {Jasniewicz}, {Jean-Antoine Piccolo}, {Jim{\'e}nez-Arranz}, {Jorissen}, {Juaristi Campillo}, {Julbe}, {Karbevska}, {Kervella}, {Khanna}, {Kontizas}, {Kordopatis}, {Korn}, {K{\'o}sp{\'a}l}, {Kostrzewa-Rutkowska}, {Kruszy{\'n}ska}, {Kun}, {Laizeau}, {Lambert}, {Lanza}, {Lasne}, {Le Campion}, {Lebreton}, {Lebzelter}, {Leccia}, {Leclerc}, {Lecoeur-Taibi}, {Liao}, {Licata}, {Lindstr{\o}m}, {Lister}, {Livanou}, {Lobel}, {Lorca}, {Loup}, {Madrero Pardo}, {Magdaleno Romeo}, {Managau}, {Mann}, {Manteiga}, {Marchant}, {Marconi}, {Marcos}, {Marcos Santos}, {Mar{\'\i}n Pina}, {Marinoni}, {Marocco}, {Marshall}, {Martin Polo}, {Mart{\'\i}n-Fleitas}, {Marton}, {Mary}, {Masip}, {Massari}, {Mastrobuono-Battisti}, {Mazeh}, {McMillan}, {Messina}, {Michalik},
  {Millar}, {Mints}, {Molina}, {Molinaro}, {Moln{\'a}r}, {Monari}, {Mongui{\'o}}, {Montegriffo}, {Montero}, {Mor}, {Mora}, {Morbidelli}, {Morel}, {Morris}, {Muraveva}, {Murphy}, {Musella}, {Nagy}, {Noval}, {Oca{\~n}a}, {Ogden}, {Ordenovic}, {Osinde}, {Pagani}, {Pagano}, {Palaversa}, {Palicio}, {Pallas-Quintela}, {Panahi}, {Payne-Wardenaar}, {Pe{\~n}alosa Esteller}, {Penttil{\"a}}, {Pichon}, {Piersimoni}, {Pineau}, {Plachy}, {Plum}, {Poggio}, {Pr{\v{s}}a}, {Pulone}, {Racero}, {Ragaini}, {Rainer}, {Raiteri}, {Rambaux}, {Ramos}, {Ramos-Lerate}, {Re Fiorentin}, {Regibo}, {Richards}, {Rios Diaz}, {Ripepi}, {Riva}, {Rix}, {Rixon}, {Robichon}, {Robin}, {Robin}, {Roelens}, {Rogues}, {Rohrbasser}, {Romero-G{\'o}mez}, {Rowell}, {Royer}, {Ruz Mieres}, {Rybicki}, {Sadowski}, {S{\'a}ez N{\'u}{\~n}ez}, {Sagrist{\`a} Sell{\'e}s}, {Sahlmann}, {Salguero}, {Samaras}, {Sanchez Gimenez}, {Sanna}, {Santove{\~n}a}, {Sarasso}, {Schultheis}, {Sciacca}, {Segol}, {Segovia}, {S{\'e}gransan}, {Semeux}, {Shahaf}, {Siddiqui}, {Siebert},
  {Siltala}, {Silvelo}, {Slezak}, {Slezak}, {Smart}, {Snaith}, {Solano}, {Solitro}, {Souami}, {Souchay}, {Spagna}, {Spina}, {Spoto}, {Steele}, {Steidelm{\"u}ller}, {Stephenson}, {S{\"u}veges}, {Surdej}, {Szabados}, {Szegedi-Elek}, {Taris}, {Taylor}, {Teixeira}, {Tolomei}, {Tonello}, {Torra}, {Torra}, {Torralba Elipe}, {Trabucchi}, {Tsounis}, {Turon}, {Ulla}, {Unger}, {Vaillant}, {van Dillen}, {van Reeven}, {Vanel}, {Vecchiato}, {Viala}, {Vicente}, {Voutsinas}, {Weiler}, {Wevers}, {Wyrzykowski}, {Yoldas}, {Yvard}, {Zhao}, {Zorec}, {Zucker}, \& {Zwitter}}]{Gaia_2023}
{Gaia Collaboration}, {Vallenari}, A., {Brown}, A.~G.~A., {et~al.} 2023, \aap, 674, A1, \dodoi{10.1051/0004-6361/202243940}

\bibitem[{{Garrido-Deutelmoser} {et~al.}(2023){Garrido-Deutelmoser}, {Petrovich}, {Charalambous}, {Guzm{\'a}n}, \& {Zhang}}]{Garrido-Deutelmoser_2023_HD163296}
{Garrido-Deutelmoser}, J., {Petrovich}, C., {Charalambous}, C., {Guzm{\'a}n}, V.~V., \& {Zhang}, K. 2023, \apjl, 945, L37, \dodoi{10.3847/2041-8213/acbea8}

\bibitem[{{Goldreich} \& {Tremaine}(1980)}]{Goldreich_Tremaine_1980}
{Goldreich}, P., \& {Tremaine}, S. 1980, \apj, 241, 425, \dodoi{10.1086/158356}

\bibitem[{{Guilloteau} {et~al.}(2014){Guilloteau}, {Simon}, {Pi{\'e}tu}, {Di Folco}, {Dutrey}, {Prato}, \& {Chapillon}}]{Guilloteau2014}
{Guilloteau}, S., {Simon}, M., {Pi{\'e}tu}, V., {et~al.} 2014, \aap, 567, A117, \dodoi{10.1051/0004-6361/201423765}

\bibitem[{{Hasegawa} {et~al.}(2017){Hasegawa}, {Okuzumi}, {Flock}, \& {Turner}}]{Hasegawa2017}
{Hasegawa}, Y., {Okuzumi}, S., {Flock}, M., \& {Turner}, N.~J. 2017, \apj, 845, 31, \dodoi{10.3847/1538-4357/aa7d55}

\bibitem[{{Hashimoto} {et~al.}(2021){Hashimoto}, {Muto}, {Dong}, {Liu}, {van der Marel}, {Francis}, {Hasegawa}, \& {Tsukagoshi}}]{Hashimoto2021}
{Hashimoto}, J., {Muto}, T., {Dong}, R., {et~al.} 2021, \apj, 911, 5, \dodoi{10.3847/1538-4357/abe59f}

\bibitem[{{Hu} {et~al.}(2024){Hu}, {Li}, {Bae}, \& {Zhu}}]{Hu2024}
{Hu}, X., {Li}, Z.-Y., {Bae}, J., \& {Zhu}, Z. 2024, arXiv e-prints, arXiv:2403.18292, \dodoi{10.48550/arXiv.2403.18292}

\bibitem[{{Huang} {et~al.}(2018){Huang}, {Andrews}, {Dullemond}, {Isella}, {P{\'e}rez}, {Guzm{\'a}n}, {{\"O}berg}, {Zhu}, {Zhang}, {Bai}, {Benisty}, {Birnstiel}, {Carpenter}, {Hughes}, {Ricci}, {Weaver}, \& {Wilner}}]{Huang2018}
{Huang}, J., {Andrews}, S.~M., {Dullemond}, C.~P., {et~al.} 2018, \apjl, 869, L42, \dodoi{10.3847/2041-8213/aaf740}

\bibitem[{{Ilee} {et~al.}(2020){Ilee}, {Hall}, {Walsh}, {Jim{\'e}nez-Serra}, {Pinte}, {Terry}, {Bourke}, \& {Hoare}}]{Ilee2020}
{Ilee}, J.~D., {Hall}, C., {Walsh}, C., {et~al.} 2020, \mnras, 498, 5116, \dodoi{10.1093/mnras/staa2699}

\bibitem[{{Jiang} \& {Ormel}(2023)}]{Jiang2023}
{Jiang}, H., \& {Ormel}, C.~W. 2023, \mnras, 518, 3877, \dodoi{10.1093/mnras/stac3275}

\bibitem[{{Kalyaan} {et~al.}(2021){Kalyaan}, {Pinilla}, {Krijt}, {Mulders}, \& {Banzatti}}]{Kalyaan2021}
{Kalyaan}, A., {Pinilla}, P., {Krijt}, S., {Mulders}, G.~D., \& {Banzatti}, A. 2021, \apj, 921, 84, \dodoi{10.3847/1538-4357/ac1e96}

\bibitem[{{Kimmig} {et~al.}(2020){Kimmig}, {Dullemond}, \& {Kley}}]{Kimmig-2020}
{Kimmig}, C.~N., {Dullemond}, C.~P., \& {Kley}, W. 2020, \aap, 633, A4, \dodoi{10.1051/0004-6361/201936412}

\bibitem[{{Krapp} {et~al.}(2018){Krapp}, {Gressel}, {Ben{\'\i}tez-Llambay}, {Downes}, {Mohandas}, \& {Pessah}}]{Krapp2018}
{Krapp}, L., {Gressel}, O., {Ben{\'\i}tez-Llambay}, P., {et~al.} 2018, \apj, 865, 105, \dodoi{10.3847/1538-4357/aadcf0}

\bibitem[{{Kudo} {et~al.}(2018){Kudo}, {Hashimoto}, {Muto}, {Liu}, {Dong}, {Hasegawa}, {Tsukagoshi}, \& {Konishi}}]{Kudo2018}
{Kudo}, T., {Hashimoto}, J., {Muto}, T., {et~al.} 2018, \apjl, 868, L5, \dodoi{10.3847/2041-8213/aaeb1c}

\bibitem[{{Lesur}(2021)}]{Lesur2021}
{Lesur}, G. R.~J. 2021, \aap, 650, A35, \dodoi{10.1051/0004-6361/202040109}

\bibitem[{{Li} {et~al.}(2001){Li}, {Colgate}, {Wendroff}, \& {Liska}}]{rwi-2001}
{Li}, H., {Colgate}, S.~A., {Wendroff}, B., \& {Liska}, R. 2001, \apj, 551, 874, \dodoi{10.1086/320241}

\bibitem[{{Lin} \& {Papaloizou}(1986)}]{Lin_Papaloizou1986a}
{Lin}, D.~N.~C., \& {Papaloizou}, J. 1986, \apj, 307, 395, \dodoi{10.1086/164426}

\bibitem[{{Lin} \& {Papaloizou}(1993)}]{Lin_Papaloizou_1993}
{Lin}, D.~N.~C., \& {Papaloizou}, J.~C.~B. 1993, in Protostars and Planets III, ed. E.~H. {Levy} \& J.~I. {Lunine}, 749--835

\bibitem[{{Lin} {et~al.}(2024){Lin}, {Wang}, {Fang}, {Nemer}, \& {Goodman}}]{Lin_Wang_2024}
{Lin}, Y., {Wang}, L., {Fang}, M., {Nemer}, A., \& {Goodman}, J. 2024, arXiv e-prints, arXiv:2401.15419, \dodoi{10.48550/arXiv.2401.15419}

\bibitem[{{Liu} \& {Bai}(2023)}]{Liu-Bai-2023}
{Liu}, H., \& {Bai}, X.-N. 2023, \mnras, 526, 80, \dodoi{10.1093/mnras/stad2629}

\bibitem[{{Liu} {et~al.}(2021){Liu}, {Tsai}, {Chen}, {Liu}, {Zhang}, {Ma}, {Elbakyan}, {Green}, {Hales}, {Liu}, {Takami}, {P{\'e}rez}, {Vorobyov}, \& {Yang}}]{Liu2021}
{Liu}, H.~B., {Tsai}, A.-L., {Chen}, W.~P., {et~al.} 2021, \apj, 923, 270, \dodoi{10.3847/1538-4357/ac31b9}

\bibitem[{{Liu} {et~al.}(2024){Liu}, {Muto}, {Konishi}, {Chung}, {Hashimoto}, {Doi}, {Dong}, {Kudo}, {Hasegawa}, {Terada}, \& {Kataoka}}]{Baobab_2024}
{Liu}, H.~B., {Muto}, T., {Konishi}, M., {et~al.} 2024, \aap, 685, A18, \dodoi{10.1051/0004-6361/202348896}

\bibitem[{{Lodato} {et~al.}(2017){Lodato}, {Scardoni}, {Manara}, \& {Testi}}]{Lodato2017}
{Lodato}, G., {Scardoni}, C.~E., {Manara}, C.~F., \& {Testi}, L. 2017, \mnras, 472, 4700, \dodoi{10.1093/mnras/stx2273}

\bibitem[{{Lodato} {et~al.}(2019){Lodato}, {Dipierro}, {Ragusa}, {Long}, {Herczeg}, {Pascucci}, {Pinilla}, {Manara}, {Tazzari}, {Liu}, {Mulders}, {Harsono}, {Boehler}, {M{\'e}nard}, {Johnstone}, {Salyk}, {van der Plas}, {Cabrit}, {Edwards}, {Fischer}, {Hendler}, {Nisini}, {Rigliaco}, {Avenhaus}, {Banzatti}, \& {Gully-Santiago}}]{lodato2019}
{Lodato}, G., {Dipierro}, G., {Ragusa}, E., {et~al.} 2019, \mnras, 486, 453, \dodoi{10.1093/mnras/stz913}

\bibitem[{{Long} {et~al.}(2018){Long}, {Pinilla}, {Herczeg}, {Harsono}, {Dipierro}, {Pascucci}, {Hendler}, {Tazzari}, {Ragusa}, {Salyk}, {Edwards}, {Lodato}, {van de Plas}, {Johnstone}, {Liu}, {Boehler}, {Cabrit}, {Manara}, {Menard}, {Mulders}, {Nisini}, {Fischer}, {Rigliaco}, {Banzatti}, {Avenhaus}, \& {Gully-Santiago}}]{long2018}
{Long}, F., {Pinilla}, P., {Herczeg}, G.~J., {et~al.} 2018, \apj, 869, 17, \dodoi{10.3847/1538-4357/aae8e1}

\bibitem[{{Long} {et~al.}(2019){Long}, {Herczeg}, {Harsono}, {Pinilla}, {Tazzari}, {Manara}, {Pascucci}, {Cabrit}, {Nisini}, {Johnstone}, {Edwards}, {Salyk}, {Menard}, {Lodato}, {Boehler}, {Mace}, {Liu}, {Mulders}, {Hendler}, {Ragusa}, {Fischer}, {Banzatti}, {Rigliaco}, {van de Plas}, {Dipierro}, {Gully-Santiago}, \& {Lopez-Valdivia}}]{Long2019}
{Long}, F., {Herczeg}, G.~J., {Harsono}, D., {et~al.} 2019, \apj, 882, 49, \dodoi{10.3847/1538-4357/ab2d2d}

\bibitem[{{Long} {et~al.}(2022){Long}, {Andrews}, {Zhang}, {Qi}, {Benisty}, {Facchini}, {Isella}, {Wilner}, {Bae}, {Huang}, {Loomis}, {{\"O}berg}, \& {Zhu}}]{Long2022LkCa15Disk}
{Long}, F., {Andrews}, S.~M., {Zhang}, S., {et~al.} 2022, \apjl, 937, L1, \dodoi{10.3847/2041-8213/ac8b10}

\bibitem[{{Lovelace} {et~al.}(1999){Lovelace}, {Li}, {Colgate}, \& {Nelson}}]{RWI-1999}
{Lovelace}, R.~V.~E., {Li}, H., {Colgate}, S.~A., \& {Nelson}, A.~F. 1999, \apj, 513, 805, \dodoi{10.1086/306900}

\bibitem[{{Luhman} {et~al.}(2010){Luhman}, {Allen}, {Espaillat}, {Hartmann}, \& {Calvet}}]{Luhman2010}
{Luhman}, K.~L., {Allen}, P.~R., {Espaillat}, C., {Hartmann}, L., \& {Calvet}, N. 2010, \apjs, 186, 111, \dodoi{10.1088/0067-0049/186/1/111}

\bibitem[{{Mathis} {et~al.}(1977){Mathis}, {Rumpl}, \& {Nordsieck}}]{MRN}
{Mathis}, J.~S., {Rumpl}, W., \& {Nordsieck}, K.~H. 1977, \apj, 217, 425, \dodoi{10.1086/155591}

\bibitem[{{McMullin} {et~al.}(2007){McMullin}, {Waters}, {Schiebel}, {Young}, \& {Golap}}]{CASA2007}
{McMullin}, J.~P., {Waters}, B., {Schiebel}, D., {Young}, W., \& {Golap}, K. 2007, in Astronomical Society of the Pacific Conference Series, Vol. 376, Astronomical Data Analysis Software and Systems XVI, ed. R.~A. {Shaw}, F.~{Hill}, \& D.~J. {Bell}, 127

\bibitem[{{McNally} {et~al.}(2019){McNally}, {Nelson}, {Paardekooper}, \& {Ben{\'\i}tez-Llambay}}]{McNally_2019}
{McNally}, C.~P., {Nelson}, R.~P., {Paardekooper}, S.-J., \& {Ben{\'\i}tez-Llambay}, P. 2019, \mnras, 484, 728, \dodoi{10.1093/mnras/stz023}

\bibitem[{{McNally} {et~al.}(2020){McNally}, {Nelson}, {Paardekooper}, {Ben{\'\i}tez-Llambay}, \& {Gressel}}]{McNally2020}
{McNally}, C.~P., {Nelson}, R.~P., {Paardekooper}, S.-J., {Ben{\'\i}tez-Llambay}, P., \& {Gressel}, O. 2020, \mnras, 493, 4382, \dodoi{10.1093/mnras/staa576}

\bibitem[{{Miranda} {et~al.}(2017){Miranda}, {Li}, {Li}, \& {Jin}}]{Miranda2017}
{Miranda}, R., {Li}, H., {Li}, S., \& {Jin}, S. 2017, \apj, 835, 118, \dodoi{10.3847/1538-4357/835/2/118}

\bibitem[{{Paardekooper} {et~al.}(2022){Paardekooper}, {Dong}, {Duffell}, {Fung}, {Masset}, {Ogilvie}, \& {Tanaka}}]{PaardekooperEtal2022}
{Paardekooper}, S.-J., {Dong}, R., {Duffell}, P., {et~al.} 2022, arXiv e-prints, arXiv:2203.09595.
\newblock \doarXiv{2203.09595}

\bibitem[{{Paardekooper} \& {Mellema}(2006)}]{paardekooper2006}
{Paardekooper}, S.~J., \& {Mellema}, G. 2006, \aap, 453, 1129, \dodoi{10.1051/0004-6361:20054449}

\bibitem[{{P{\'e}rez} {et~al.}(2019){P{\'e}rez}, {Casassus}, {Baruteau}, {Dong}, {Hales}, \& {Cieza}}]{Perez_2019}
{P{\'e}rez}, S., {Casassus}, S., {Baruteau}, C., {et~al.} 2019, \aj, 158, 15, \dodoi{10.3847/1538-3881/ab1f88}

\bibitem[{{Pi{\'e}tu} {et~al.}(2007){Pi{\'e}tu}, {Dutrey}, \& {Guilloteau}}]{DMTau145pc_2007}
{Pi{\'e}tu}, V., {Dutrey}, A., \& {Guilloteau}, S. 2007, \aap, 467, 163, \dodoi{10.1051/0004-6361:20066537}

\bibitem[{{Pinilla} {et~al.}(2012){Pinilla}, {Benisty}, \& {Birnstiel}}]{Pinilla_2012}
{Pinilla}, P., {Benisty}, M., \& {Birnstiel}, T. 2012, \aap, 545, A81, \dodoi{10.1051/0004-6361/201219315}

\bibitem[{{Pinilla} {et~al.}(2021){Pinilla}, {Lenz}, \& {Stammler}}]{Pinilla_2021}
{Pinilla}, P., {Lenz}, C.~T., \& {Stammler}, S.~M. 2021, \aap, 645, A70, \dodoi{10.1051/0004-6361/202038920}

\bibitem[{{Ricci} {et~al.}(2021){Ricci}, {Harter}, {Ercolano}, \& {Weber}}]{Ricci2021}
{Ricci}, L., {Harter}, S.~K., {Ercolano}, B., \& {Weber}, M. 2021, \apj, 913, 122, \dodoi{10.3847/1538-4357/abf5d8}

\bibitem[{{Ricci} {et~al.}(2018){Ricci}, {Liu}, {Isella}, \& {Li}}]{Ricci2018}
{Ricci}, L., {Liu}, S.-F., {Isella}, A., \& {Li}, H. 2018, \apj, 853, 110, \dodoi{10.3847/1538-4357/aaa546}

\bibitem[{{Rosotti} {et~al.}(2016){Rosotti}, {Juhasz}, {Booth}, \& {Clarke}}]{Rosotti2016}
{Rosotti}, G.~P., {Juhasz}, A., {Booth}, R.~A., \& {Clarke}, C.~J. 2016, \mnras, 459, 2790, \dodoi{10.1093/mnras/stw691}

\bibitem[{{Rosotti} {et~al.}(2020){Rosotti}, {Teague}, {Dullemond}, {Booth}, \& {Clarke}}]{Rosotti2020}
{Rosotti}, G.~P., {Teague}, R., {Dullemond}, C., {Booth}, R.~A., \& {Clarke}, C.~J. 2020, \mnras, 495, 173, \dodoi{10.1093/mnras/staa1170}

\bibitem[{{Sellek} {et~al.}(2024){Sellek}, {Bajaj}, {Pascucci}, {Clarke}, {Alexander}, {Xie}, {Ballabio}, {Deng}, {Gorti}, {Gaspar}, \& {Morrison}}]{Sellek2024}
{Sellek}, A.~D., {Bajaj}, N.~S., {Pascucci}, I., {et~al.} 2024, arXiv e-prints, arXiv:2403.09780, \dodoi{10.48550/arXiv.2403.09780}

\bibitem[{{Shakura} \& {Sunyaev}(1973)}]{Shakura-Sunyaev73}
{Shakura}, N.~I., \& {Sunyaev}, R.~A. 1973, \aap, 24, 337

\bibitem[{{Simon} {et~al.}(2015){Simon}, {Hughes}, {Flaherty}, {Bai}, \& {Armitage}}]{Simon2015}
{Simon}, J.~B., {Hughes}, A.~M., {Flaherty}, K.~M., {Bai}, X.-N., \& {Armitage}, P.~J. 2015, \apj, 808, 180, \dodoi{10.1088/0004-637X/808/2/180}

\bibitem[{{Simon} {et~al.}(2000){Simon}, {Dutrey}, \& {Guilloteau}}]{Simon2000}
{Simon}, M., {Dutrey}, A., \& {Guilloteau}, S. 2000, \apj, 545, 1034, \dodoi{10.1086/317838}

\bibitem[{{Stadler} {et~al.}(2022){Stadler}, {G{\'a}rate}, {Pinilla}, {Lenz}, {Dullemond}, {Birnstiel}, \& {Stammler}}]{Stadler_2022}
{Stadler}, J., {G{\'a}rate}, M., {Pinilla}, P., {et~al.} 2022, \aap, 668, A104, \dodoi{10.1051/0004-6361/202243338}

\bibitem[{{Suzuki} {et~al.}(2016){Suzuki}, {Ogihara}, {Morbidelli}, {Crida}, \& {Guillot}}]{Suzuki2016}
{Suzuki}, T.~K., {Ogihara}, M., {Morbidelli}, A., {Crida}, A., \& {Guillot}, T. 2016, \aap, 596, A74, \dodoi{10.1051/0004-6361/201628955}

\bibitem[{{Tabone} {et~al.}(2022{\natexlab{a}}){Tabone}, {Rosotti}, {Cridland}, {Armitage}, \& {Lodato}}]{Tabone22}
{Tabone}, B., {Rosotti}, G.~P., {Cridland}, A.~J., {Armitage}, P.~J., \& {Lodato}, G. 2022{\natexlab{a}}, \mnras, 512, 2290, \dodoi{10.1093/mnras/stab3442}

\bibitem[{{Tabone} {et~al.}(2022{\natexlab{b}}){Tabone}, {Rosotti}, {Lodato}, {Armitage}, {Cridland}, \& {van Dishoeck}}]{Tabone2022-2}
{Tabone}, B., {Rosotti}, G.~P., {Lodato}, G., {et~al.} 2022{\natexlab{b}}, \mnras, 512, L74, \dodoi{10.1093/mnrasl/slab124}

\bibitem[{{Trapman} {et~al.}(2022){Trapman}, {Zhang}, {van't Hoff}, {Hogerheijde}, \& {Bergin}}]{Trapman_2022}
{Trapman}, L., {Zhang}, K., {van't Hoff}, M. L.~R., {Hogerheijde}, M.~R., \& {Bergin}, E.~A. 2022, \apjl, 926, L2, \dodoi{10.3847/2041-8213/ac4f47}

\bibitem[{{Turpin} \& {Nelson}(2024)}]{Turpin2024}
{Turpin}, G.~A., \& {Nelson}, R.~P. 2024, \mnras, 528, 7256, \dodoi{10.1093/mnras/stae109}

\bibitem[{{Wu} {et~al.}(2023{\natexlab{a}}){Wu}, {Baruteau}, \& {Nayakshin}}]{Wu_2023_migration}
{Wu}, Y., {Baruteau}, C., \& {Nayakshin}, S. 2023{\natexlab{a}}, \mnras, 523, 4869, \dodoi{10.1093/mnras/stad1791}

\bibitem[{{Wu} {et~al.}(2023{\natexlab{b}}){Wu}, {Chen}, {Jiang}, {Dong}, {Mac{\'\i}as}, {Lin}, {Rosotti}, \& {Elbakyan}}]{WU_Chen_jiang_2023}
{Wu}, Y., {Chen}, Y.-X., {Jiang}, H., {et~al.} 2023{\natexlab{b}}, \mnras, 523, 2630, \dodoi{10.1093/mnras/stad1553}

\bibitem[{{Wu} {et~al.}(2024{\natexlab{a}}){Wu}, {Chen}, \& {Lin}}]{Wu_2024_chaotic}
{Wu}, Y., {Chen}, Y.-X., \& {Lin}, D. N.~C. 2024{\natexlab{a}}, \mnras, 528, L127, \dodoi{10.1093/mnrasl/slad183}

\bibitem[{{Wu} {et~al.}(2024{\natexlab{b}}){Wu}, {Lin}, {Cui}, {Krapp}, {Lee}, \& {Youdin}}]{Wu_2024_BDHI}
{Wu}, Y., {Lin}, M.-K., {Cui}, C., {et~al.} 2024{\natexlab{b}}, \apj, 962, 173, \dodoi{10.3847/1538-4357/ad15fe}

\bibitem[{{Wu} {et~al.}(2024{\natexlab{c}}){Wu}, {Liu}, {Jiang}, \& {Nayakshin}}]{Wu_2024_ska_ngvla}
{Wu}, Y., {Liu}, S.-F., {Jiang}, H., \& {Nayakshin}, S. 2024{\natexlab{c}}, \apj, 965, 110, \dodoi{10.3847/1538-4357/ad323b}

\bibitem[{{Xu}(2022)}]{Xu2022}
{Xu}, W. 2022, \apj, 934, 156, \dodoi{10.3847/1538-4357/ac7b94}

\bibitem[{{Xu} {et~al.}(2023){Xu}, {Ohashi}, {Aso}, \& {Liu}}]{Xu2023}
{Xu}, W., {Ohashi}, S., {Aso}, Y., \& {Liu}, H.~B. 2023, \apj, 954, 190, \dodoi{10.3847/1538-4357/aced4c}

\bibitem[{{Xu} \& {Wang}(2024)}]{Xu-Wang-2024}
{Xu}, W., \& {Wang}, S. 2024, \apjl, 962, L4, \dodoi{10.3847/2041-8213/ad1ee1}

\bibitem[{{Youdin} \& {Lithwick}(2007)}]{Youdin-Lithwick}
{Youdin}, A.~N., \& {Lithwick}, Y. 2007, \icarus, 192, 588, \dodoi{10.1016/j.icarus.2007.07.012}

\bibitem[{{Zapata} {et~al.}(2017){Zapata}, {Rodr{\'\i}guez}, \& {Palau}}]{Zapata2017}
{Zapata}, L.~A., {Rodr{\'\i}guez}, L.~F., \& {Palau}, A. 2017, \apj, 834, 138, \dodoi{10.3847/1538-4357/834/2/138}

\bibitem[{{Zhang} {et~al.}(2018){Zhang}, {Zhu}, {Huang}, {Guzm{\'a}n}, {Andrews}, {Birnstiel}, {Dullemond}, {Carpenter}, {Isella}, {P{\'e}rez}, {Benisty}, {Wilner}, {Baruteau}, {Bai}, \& {Ricci}}]{2018ZhangDSHARP}
{Zhang}, S., {Zhu}, Z., {Huang}, J., {et~al.} 2018, \apjl, 869, L47, \dodoi{10.3847/2041-8213/aaf744}

\bibitem[{{Zhou} {et~al.}(2022){Zhou}, {Deng}, {Chen}, \& {Lin}}]{Zhou2022}
{Zhou}, T., {Deng}, H.-P., {Chen}, Y.-X., \& {Lin}, D. N.~C. 2022, \apj, 940, 117, \dodoi{10.3847/1538-4357/ac9bf6}

\bibitem[{{Zhu} {et~al.}(2011){Zhu}, {Nelson}, {Hartmann}, {Espaillat}, \& {Calvet}}]{Zhuetal2011}
{Zhu}, Z., {Nelson}, R.~P., {Hartmann}, L., {Espaillat}, C., \& {Calvet}, N. 2011, \apj, 729, 47, \dodoi{10.1088/0004-637X/729/1/47}

\bibitem[{{Zhu} {et~al.}(2015){Zhu}, {Stone}, \& {Bai}}]{Zhu-2015}
{Zhu}, Z., {Stone}, J.~M., \& {Bai}, X.-N. 2015, \apj, 801, 81, \dodoi{10.1088/0004-637X/801/2/81}

\bibitem[{{Ziampras} {et~al.}(2024){Ziampras}, {Nelson}, \& {Paardekooper}}]{Ziampras2024}
{Ziampras}, A., {Nelson}, R.~P., \& {Paardekooper}, S.-J. 2024, \mnras, 528, 6130, \dodoi{10.1093/mnras/stae372}

\end{thebibliography}
\bibliographystyle{aasjournal}

\end{CJK*}
\end{document}